\documentclass[10pt,twocolumn]{article}
\pdfoutput=1
\usepackage{times,ifpdf}
\usepackage{fullpage}
\usepackage{authblk}
\usepackage{color,courier}
\usepackage[normalem]{ulem}
\usepackage{multirow}
\usepackage{listings}
\definecolor{gray}{rgb}{0.5,0.5,0.5}
\definecolor{mauve}{rgb}{0.25,0.50,0.50}
\usepackage{tikz}
\usetikzlibrary{calc,positioning}
\usepackage{colortbl}
\usepackage{makecell}
\usepackage{comment}
\usepackage[caption=false]{subfig}
\definecolor{dkgreen}{rgb}{0,0.6,0}
\definecolor{gray}{rgb}{0.5,0.5,0.5}
\usepackage{epstopdf}
\usepackage{amsmath,amsthm}
\usepackage{amssymb}
\usepackage{stmaryrd}
\usepackage{algorithmic}
\usepackage[ruled,vlined,linesnumbered,nofillcomment]{algorithm2e}
\usepackage{caption}
\newtheorem{mydef}{RULE}
\newtheorem{mydef1}{Theorem}

\usepackage[english]{babel}
\makeatletter
\adddialect\l@ENGLISH\l@english
\makeatother

\newcommand{\minitab}[2][l]{\begin{tabular}{#1}#2\end{tabular}}
\newcommand{\tikzmark}[1]{\tikz[overlay,remember picture] \node (#1) {};}

\addto\captionsenglish{}
\addto\captionsenglish{}

\begin{document}





\title{\textbf{On Performance Debugging of Unnecessary Lock Contentions on Multicore Processors: A Replay-based Approach}}
\author[1]{Long Zheng}
\author[1]{Xiaofei Liao}
\author[2]{Bingsheng He}
\author[1]{Song Wu}
\author[1]{Hai Jin}
\affil[1]{\em Huazhong University of Science and Technology}
\affil[2]{\em Nanyang Technological University}

\date{}
\maketitle
\thispagestyle{empty}
\begin{abstract}
Locks have been widely used as an effective synchronization mechanism among processes and threads. However, we observe that, a large number of false inter-thread dependencies (i.e., unnecessary lock contention) exist during the execution on multicore processors, which incurs significant performance overhead. This paper, therefore, presents a performance debugging framework, \textsc{PerfPlay}, to facilitate a comprehensive and in-depth \emph{understanding} of the performance impact of unnecessary lock contentions. The core technique of our debugging framework is trace replay. Specifically, \textsc{PerfPlay} records the program execution trace, on the basis of which we can detect all unnecessary lock contentions through trace replay techniques. We propose a novel technique of trace transformation to transform these unnecessary lock contentions in the original trace into the correct pattern as a new trace free from unnecessary lock contentions. Through replaying both traces, \textsc{PerfPlay} can evaluate the performance impact of each unnecessary lock contention. To demonstrate the effectiveness of \textsc{PerfPlay}, we study five real world programs and PARSEC benchmarks. Our experimental results demonstrate the significant performance overhead of unnecessary lock contentions, and the effectiveness of our performance debugging framework in identifying the performance critical unnecessary lock contentions in real applications.
\end{abstract}


\section{Introduction}
\label{sec:intro}
\begin{figure}[t]
\centering
\begin{tabular}{c}
\begin{lstlisting}[escapechar=@,linewidth=7.8cm]
Thread 1:
void fil_flush_file_spaces(...){
5609:   @\textbf{mutex\_enter}@(&fil_system->mutex);
5611:   n_space_ids=UT_LIST_GET_LEN(
                 @\framebox[1\width]{\textbf{fil->system->unflushed\_space$\tikzmark{L1line1}$s}}@);
5614:   @\textbf{mutex\_exit}@(&fil_system->mutex);
}
Thread 2:                            @\textsf{\textbf{Real data conflict}}@
void fil_flush(...){
5473:   @\textbf{mutex\_enter}@(&fil_system->mutex);
        //search hash table by a given id
5475:   space=fil_space_get_by_id(space_id);
5483:   if (fil_buffering_disabled(space)) {
            //checking some data and states
5501:       @\textbf{mutex\_exit}@(&fil_system->mutex);
5503:       return;}
            ......
5573:   UT_LIST_REMOVE(unflushed_spaces,
            @\framebox[1\width]{\textbf{fil->system->unflushed\_space$\tikzmark{L2line1}$s}}@,space);
5592:   @\textbf{mutex\_exit}@(&fil_system->mutex);
}                             @\textsf{\textit{storage/innobase/fil/fil0fil.cc}}@
\end{lstlisting}
\tikz[overlay,remember picture,<->] \draw[color=black,thick,out=-70,in=40] ($(L1line1)+(0ex,-0.5ex)$) to ($(L2line1)+(0pt,0.9ex)$) ;
\end{tabular}
\caption{An example of the potential parallelism serialized by the unnecessary lock contention from \texttt{mysql} in the dynamic execution}
\label{fig:1}
\end{figure}
In the era of multi-core processors, parallel programming is prevalent. The efficiency of process/thread communication is very important for the overall performance of parallel executions. In multi-threaded applications, locks are widely-used to ensure the mutual access to shared data within the critical sections. 
However, during the dynamic execution on multicore processors, multiple critical sections protected by the same lock do not necessarily conflict at runtime. 
Therefore, a program may produce false inter-thread dependency (i.e., unnecessary lock contention). Such unnecessary lock contentions serialize the access, thus leading to the severe performance loss of programs~\cite{SLE, TLE}. 

Let us illustrate the problem of unnecessary lock contentions with a real example. Figure~\ref{fig:1} is an example from \textsf{mysql}$-5.6.11$, a database object repository \cite{mysql}, which depicts how the unnecessary lock contention occur in the dynamic execution. Both threads hold the same shared lock \texttt{fil\_system->mutex} to coordinate the shared access to \texttt{fil->system->unflushed->spaces}. However, in the dynamic execution, the thread always does not update it, if the buffer is explicitly disabled by the user (i.e., \texttt{fil\_buffering\_disabled(space)}={\sf true}). In this case, two threads do not conflict, and the lock unnecessarily serializes the operations of \texttt{UT\_LIST\_GET\_LEN} and \texttt{fil\_space\_get\_by\_id}, thus leading to the performance degradation.
In practice, we need to identify and generalize \emph{Unnecessary Lock Contention Pair} (ULCP), which consists of two critical sections protected by the same lock accessing the \emph{parallelizable} regions.

Due to the significant overhead of ULCPs at runtime,
a volume of research~\cite{SLE, TLE, Roy:2009} attempts to dynamically eliminate the performance impact of ULCPs by speculatively executing critical sections without actually acquiring the lock. The lock is taken only when a data conflict needs to be resolved.
The major advantage of those approaches is that they are transparent to programmers. However, they incur many problems in practice~\cite{Leis:2014, Yehuda:2014} and are still a challenge and long way before their practical and wide adoptions. First, they are prone to trigger false aborts due to the hardware limitations \cite{Leis:2014}.
Second, a few transaction aborts (including false aborts) may cause overmany rollbacks~\cite{Yehuda:2014}. As a result, a volume of extra runtime overhead can be reintroduced.
Instead of relying on complicated dynamic approaches, this paper argues that the programmer should play a proactive role in eliminating the overhead of ULCPs. If the programmer can fix the performance problem caused by ULCPs, the side-effect problems of existing ULCP tools~\cite{SLE, TLE, Roy:2009, Haswell, Cain:2013} can be avoided. We study five real-world programs and PARSEC benchmarks to probe into the explicit characteristics of ULCPs. Based on our observations, we get an important finding: the root cause of many ULCPs lies in the problematic synchronization implementation. These ULCPs can be fixed by programmers. Thus, it is necessary to detect them and further assist the programmers to understand and correct them, rather than take tolerant attitudes towards them in the previous work. However, it is a nontrivial task to identify the source of ULCPs as well as to figure out their performance impact. In fact, in a multi-threaded program, there may be so many ULCPs that it is difficult, or even impossible, to check all the ULCPs manually. Even worse, they are interwined with each other in the source code.

To help the programmer address the ULCP problems, this paper presents a performance debugging framework (namely \textsc{PerfPlay}) to \emph{understand} the performance impact of ULCPs in the lock-based programs. The core idea of \textsc{PerfPlay} is based on record/replay. Under this framework, the ULCP analysis is performed as the following steps.
\textsc{PerfPlay} first records program execution into a trace.
Through analyzing the original trace, \textsc{PerfPlay} can identify all ULCPs in the original execution. Then we propose a
novel technique of trace transformation formalized by four rules to transform these ULCPs in the original trace into the correct pattern as a new trace free from ULCPs. We ensure that the new ULCP-free trace can be performed with the correct program semantics in the most cases. 
By replaying the original trace and ULCP-free one, \textsc{PerfPlay} gets the performance impact of each ULCP. Finally we group the ULCPs generated by the same code regions and summarize the overall performance per code-site. As a result, we can recommend the programmer to fix the identified code region that would incur the highest performance impact.

Our experimental results on five real programs and PARSEC benchmarks demonstrate the performance fidelity (including performance stability and precision) and the efficiency ($<4.3\%$ lockset overhead) of \textsc{PerfPlay}. With the most beneficial code regions recommended by \textsc{PerfPlay}, our case studies verify the effectiveness of \textsc{PerfPlay} in identifying the performance critical ULCPs, and it is also revealed that the majority of ULCPs can be resolved by taking the code regions
that are most beneficial to optimize.

The rest of this paper proceeds as follows. We provide the introduction on ULCP, the motivation and overview of our work in Section~\ref{sec:background}. Section~\ref{sec:ILCP} elaborates how to transform a recorded ULCP trace into a new ULCP-free trace. Section~\ref{sec:replay} describes how to assess the ULCP performance impact from two replayed results. Section~\ref{sec:perfplay} further presents the implementation details. Section~\ref{sec:evaluation} presents the experimental results. We review the related work in Section~\ref{sec:relatedwork} and Section~\ref{sec:conclusion} concludes the work.

\section{ULCP: The Classification, A Brief Study, and Motivation}
\label{sec:background}
We start with a motivation study on ULCPs. We have observed that ULCPs are a very common problem in many multi-threaded programs. Next, we give another concrete example to show the performance impact of ULCPs. Motivated by the study and example, we develop a debugging framework to address the ULCP problem.
\begin{table}[t]   
\setlength{\abovecaptionskip}{10pt}
\setlength{\belowcaptionskip}{-10pt}
\centering
\scriptsize
\tabcolsep=0.1cm
\caption{Breakdown of ULCPs in real-world programs and PARSEC benchmarks. $Size$ is denoted to the code size of programs, $\#Locks$ the number of lock protections generated in dynamic execution. $NL.$ refers to the null-locks, $RR.$ the read-read pattern, $DW.$ the pattern of disjoint-write.}
\label{table:all}
\begin{tabular}{|c|r|r|r|r|r|r|r|}\hline
  \multirow{2}{*}{\textbf{Applications}} &\multirow{2}{*}{\textbf{LOC}}&\multirow{2}{*}{\textbf{Size}}& \multirow{2}{*}{\textbf{\# Locks}}&\multicolumn{4}{c|}{\textbf{\# ULCPs}}\\
  \cline{5-8}
  & & & &\textbf{NL.} &\textbf{RR.} &\textbf{DW.}&\textbf{Bengin} \\ \hline \hline
  {\sf openldap}        &392K&6M&1,851&75&1,414&473&15 \\
  {\sf mysql}         &1,132K&22M&2,109&125&9,822&2,924&194 \\ \hline
  {\sf pbzip2}        &5K&1M&1,281&2&1047&838&51 \\
  {\sf transmissionBT}        &79K&4M&352&15&111&123&29 \\
  {\sf handbrake}        &1,070K&3M&18,316&10&1,536&1,143&189 \\ \hline
  {\sf blackscholes}          &812&204K&0&0&0&0&0 \\
  {\sf bodytrack}          &10K&9.0M&32,642&0&1,322&321&43 \\
  {\sf canneal}       &4K&628K&34&0&0&0&0 \\
  {\sf dedup}         &3.6K&156K&19,352&231&2,421&1,952&164 \\
  {\sf facesim}       &29K&4.8K&14,541&102&871&819&12 \\
  {\sf ferret}        &9.7K&316K&6,231&11&101&231&343 \\
  {\sf fluidanimate}          &1.4K&72K&82,142&2&10,501&6,694&197 \\
  {\sf streamcluster}          &1.3K&44K&191&0&0&0&0 \\
  {\sf swaptions}          &1.5K&152K&23&0&0&0&0 \\
  {\sf vips}          &3.2K&17M&33,586&142&4,512&1,142&26 \\
  {\sf x264}          &40.3K&2.4M&16,767&941&3,841&412&84 \\ \hline
\end{tabular}
\end{table}
\subsection{A Motivation Study}
\label{ULCP:trans:classification}

\begin{figure}[t] 
\centering
\includegraphics[scale=1.7]{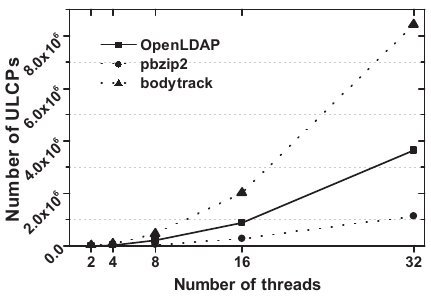}
\caption{Number of ULCPs with the increasing number of threads}
\label{fig:growth}
\end{figure}
\begin{figure}[t]
\centering
\begin{tabular}{c}
\begin{lstlisting}[escapechar=@]
@\textbf{Lock(L)}@
if(local_variable)
    shared_variable++;
@\textbf{Unlock(L)}@
\end{lstlisting}
\end{tabular}
\caption{A generic model for the generation of \texttt{null-lock}. If \texttt{local\_variable} is always read with \texttt{false} by many threads, a large number of \texttt{null-locks} will occur.}
\label{fig:nulllock}
\end{figure}
\begin{figure*}[tbp]
\centering
\includegraphics[scale=0.7]{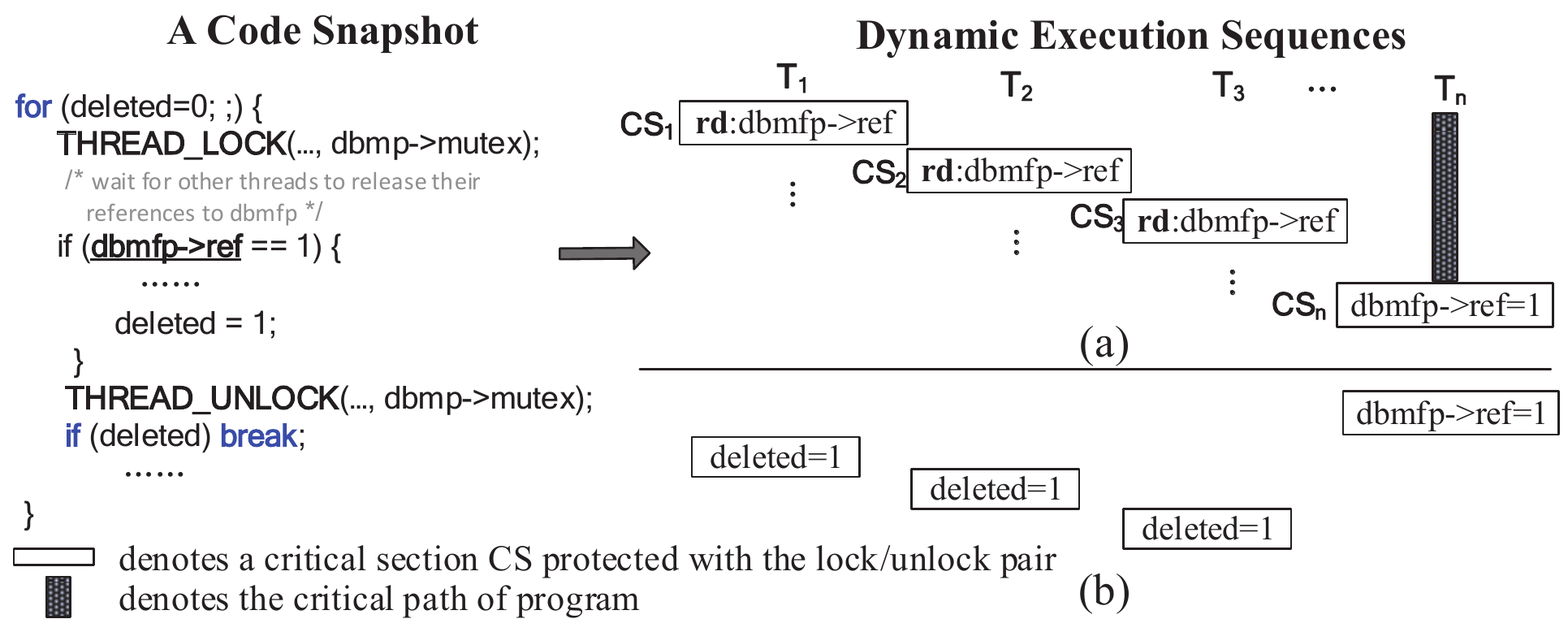}
\caption{A code snippet with problematic synchronization implementation from \textsf{OpenLDAP} and its possible dynamic execution sequences when many threads call this code simultaneously. (a) A great deal of CPU time is wasted due to the spin-waits of threads $T_0,\cdots,T_{n-1}$ for the release of \texttt{dbmfp->ref} if the critical thread $T_n$ runs slowly. (b) Little CPU time is wasted if $T_n$ is finished fast.}
\label{fig:mtexample}
\end{figure*}

We have surveyed the number of each category of ULCPs in five real-world programs (including two servers \emph{openldap}~\cite{OpenLDAP}, \emph{mysql}~\cite{mysql}; three desktop applications \emph{pbzip2}~\cite{pbzip2}, \emph{transmissionBT}~\cite{transmissionbt} and \emph{handBrake}~\cite{handbrake}) and PARSEC benchmarks~\cite{parsec}.

In our study, {we consider the ULCP in the format of pairs, because pairs are the most basic representation and can be used to represent other complex combinations beyond pairs. For instance, three sequential critical sections can be encoded as two pairs. }Our study has observed the following four major kinds of ULCPs.

(1) \emph{Null-Lock} refers to the synchronization pair where there exists no shared-memory access in the critical sections. ULCP problems of this type are usually relatively easy to understand and identify. Null-locks usually come from \emph{if-branch} of the program~\cite{ICR}. In Figure~\ref{fig:nulllock}, if \texttt{local\_variable} is accessed by multiple threads with the value of \textsf{false}, \texttt{shared\_variable} is not accessed. In this case, the program produces a large number of null-locks. We can fix null lock of this example by moving the lock and unlock operations into the scope of if-branch.

(2) \emph{Read-Read} pattern indicates that only read operations on shared data exist between two critical sections protected by the same lock. The performance problem of this type mainly stems from the serial access to the shared memory location, especially for those memory intensive applications. Figure~\ref{fig:mtexample} demonstrates such a ULCP problem from {\sf OpenLDAP}~\cite{OpenLDAP}.

(3) \emph{Disjoint-Write} pattern occurs in the scenario where two critical sections protected by the same lock update different shared memory locations, and at least one of them is the write operation. One common example of disjoint-write is that program uses the uniform reference (e.g., pointer alias) protected by the same lock to update different shared objects.

(4) \emph{Benign} pattern represents the benign feature of some \emph{false} conflicting ULCPs. To be specific, two critical sections indeed access the same shared data concurrently but they do not constitute a conflicting pair, such as redundant writes, disjoint bit manipulation, and ad-hoc synchronization \cite{Narayanasamy2007, DataCollider}.

Table~\ref{table:all} lists the quantitative distribution of ULCPs of all applications with two threads.
According to Table~\ref{table:all}, we find that ULCPs are pervasive. Meanwhile, different applications generally show different characteristics of ULCPs.
Figure~\ref{fig:growth} shows the growth trend of ULCPs in \emph{openldap}, \emph{pbzip2}, \emph{bodytrack} as the number of threads increases. In Figure~\ref{fig:growth}, all applications have a quantity-increasing problem, almost close to proportional order to the number of threads. This phenomenon emerges due to the reason that the occurrence of ULCPs, in most cases, is interconnected rather than isolated. The ULCP interconnection can be embodied in the fact that they are produced by some common codes that will be repeatedly executed in most threads.

The four classified categories of ULCPs facilitates the achievement of two goals: 1) ULCP identification: different patterns may involve different detection techniques; and 2) ULCP transformation (i.e., trace-level ULCP elimination): after ULCP identification, we need to transform these ULCPs into the parallel-style in the trace, but different patterns may require different fix strategies.

\subsection{Another Motivating Example}

Figure~\ref{fig:mtexample} depicts a source code snippet protected by the lock \texttt{dbmp->mutex} from \textsf{OpenLDAP}, a lightweight directory access protocol server~\cite{OpenLDAP}. This piece of code may affect the CPU utilization of system when a large number of threads call this code simultaneously. That is because it produces a large number of lock/unlock pairs (i.e., critical sections, $CS$s) where no effective execution statement exists if \texttt{dbmfp->ref} is always read by \textsf{false}. In fact, these shared reads can be operated simultaneously before \texttt{dbmfp->ref} is assigned with \textsf{true}. Figure~\ref{fig:mtexample}(a) illustrates many ULCPs (i.e., a two-tuple consisting of two critical sections $\langle CS, CS\rangle$), such as $\langle CS_1, CS_2\rangle$ and $\langle CS_2, CS_3\rangle$. ULCP introduces subtle performance impact due to the lock protection serializing two critical sections. However, if we group each ULCP based on its code-site, ULCP group would accumulate a profitable performance gain. For instance, $\langle CS_1, CS_2\rangle$ and $\langle CS_2, CS_3\rangle$ are both generated by the pair of above-depicted source code, therefore their performance benefits should be accumulated up when we evaluate the ULCP performance impact per code-site.

Lock Elision (LE)~\cite{SLE} is a technique that dynamically eliminates the inter-thread ULCP dependencies. Previous studies based on LE~\cite{SLE, TLE, Roy:2009, Yehuda:2014} resolve ULCPs of the parallel execution at runtime, which do not offer debugging information to programmers. For the example in Figure~\ref{fig:mtexample}, they remove the lock acquisition and release operations of the critical sections (i.e., $CS_1,\cdots,CS_{n-1}$) completely before $CS_n$ is executed.  As a result, $CS_1,\cdots,CS_{n-1}$ are performed in parallel. LE cannot precisely track the impact of system resource waste for ULCPs. The root cause of the problem in this example can be attributed to the imperfect synchronization implementation. To understand and fix this problem, it is necessary to detect the code regions producing ULCPs for the programmers and help the programmers further understand and correct them. In fact, this source snapshot performs the same function as \textsf{barrier} primitive. Programmers can use \textsf{barrier} primitive to fix the problem and obtain better CPU utilization.
\begin{figure*}[tbp] 
\centering
\includegraphics[scale=1]{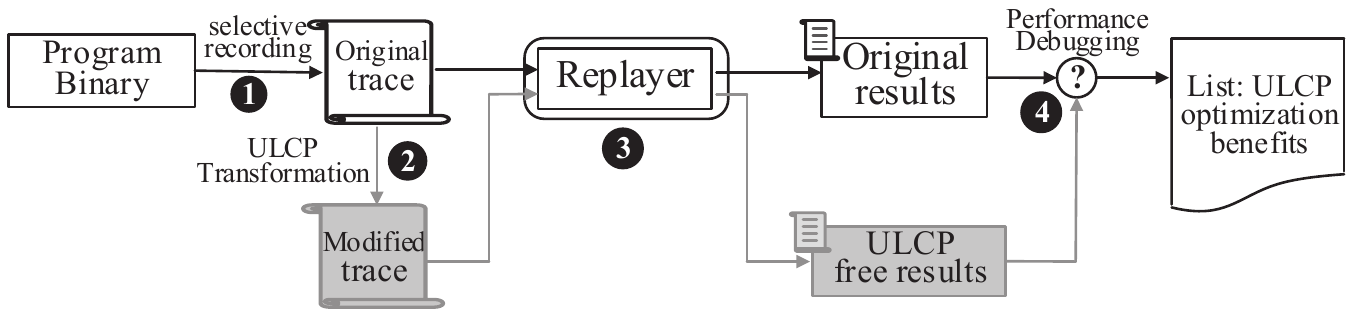}
\caption{Overview of \textsc{PerfPlay}}
\label{fig:overview}
\end{figure*}
\subsection{Overview of Our Approach}
From the aforementioned study and real-world example, we can at least get two important implications:
\begin{itemize}
  \item ULCP is a diverse program behavior. It is ubiquitous in the multi-threaded program and scattered in the program execution;
  \item It is difficult, or even impossible, to manually figure out which code-site incurs the highest performance impact due to ULCPs.
\end{itemize}

Therefore, a performance debugging tool is needed to assist the programmer in addressing the problem of ULCPs in their code. Particularly, we propose \textsc{PerfPlay}, a replay framework to help the programmers understand ULCPs in two aspects.
First, replay system records the program execution into a trace, based on which we therefore can know the explicit characteristic of each ULCP and further aggregate them according to their code-site.
Second, replay system provides the possibility of reproducing the program execution, so that we can assess the performance impact of ULCPs for the performance comparison before and after optimization to further determine the most beneficial one.

Figure~\ref{fig:overview} depicts the overview of \textsc{PerfPlay}. \textsc{PerfPlay} operates entirely on application binaries without any source code, and reports a list of the potential optimization benefits. This list is used to assist programmers to understand the ULCP performance problems.
The first step of \textsc{PerfPlay} is to record the intervals of a program execution trace. 
After the generation of original recording trace, the second step of \textsc{PerfPlay} is to transform the original trace with ULCPs into a new trace without ULCPs. Next, \textsc{PerfPlay} replays the original trace and the modified one without ULCPs.
By comparing these two replayed results before and after optimization, \textsc{PerfPlay} finally evaluates the potential performance impact of the aggregated ULCPs per code-site.

Using record/reply as the key technique for addressing ULCPs, we have addressed the following two major challenges. First, the ULCP transformation may change the synchronization structure of program, thus possibly incurring the incorrect program semantics. There lacks a mechanisms in record/replay to ensure program correctness. \textsc{PerfPlay} develops novel rule-based trace transformation techniques to preserve the program semantics (Section~\ref{sec:ILCP}). Second, we need to assess the performance of a ULCP, and determine the code-site which produces the highest performance impact due to ULCPs. PERFPLAY relies on replaying the original trace and the trace after eliminating ULCPs~(Section~\ref{sec:replay}).


\section{ULCP Transformation}
\label{sec:ILCP}
\begin{figure*}[t]
\centering
\includegraphics[scale=1.4]{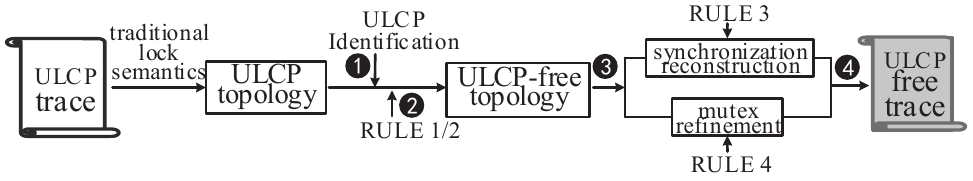}
\caption{The process of ULCP transformation}
\label{fig:transfprocess}
\end{figure*}
This section presents the detailed procedure of transforming the original trace with ULCPs into a new trace without ULCPs. The ULCP transformation may involve a change of the synchronization structure, thus making it a major threat to the program semantics. To cope with this problem, we propose a novel technique of trace transformation. We model the trace transformation problem into the graph analysis by means of topological graph theory \cite{TGT}. Since topological graph theory has been studied for decades, the ULCP problem can be solved easily by analyzing the graph.

The basic idea is as follows. We first build a topological graph which contains the original ULCP problems. Through some technical graph analyses, we then can easily identify the ULCPs and further eliminate them based on this graph as a new topological graph exclusive of ULCPs. As the topological graph can not be recognized to perform a program execution by computers, it is necessary to re-construct the ULCP-free program structure the new topological graph represents so that the computer can perform the new ULCP-free program execution.
To facilitate the description, we make the following definitions:
\begin{itemize}
 \item \textbf{\emph{Causal-order topology}:} a topological graph of the cause and effect of an execution trace. If there is no special instruction, the causal-order topology can be also referred to as topology for short.
 \item \textbf{\emph{Node}:} a critical section in the topology.
 \item \textbf{\emph{Causal-edge}:} a specific causality between nodes.
\end{itemize}

Figure~\ref{fig:transfprocess} depicts the detailed process of our trace transformation. It is a rule-based approach.

\begin{itemize}
  \item RULE 1: Following the traditional lock dependencies, we first build the causal-order topology of original execution (\emph{abbr}. original topology). The original topology involves many causal-edges caused by ULCPs. Thus we then transform original ULCP topology into a new topology which does not contain causal-edges caused by ULCPs (\emph{abbr}. ULCP-free topology).
  \item RULE 2: To ensure the stable performance for the performance analysis, it is necessary to ensure the partial orders of all causal-edge nodes in the given ULCP-free topology.
  \item RULE 3: Based on the given ULCP-free topology, we re-establish the program structure of ULCP-free topology with the help of lockset~\cite{Eraser}.
  \item RULE 4: We further refine the mutex relation for the ULCP-free trace execution.
\end{itemize}

Based on the four rules proposed, the new ULCP-free trace is performed with the correct program semantics in most cases. If not, it would report the data races.
Next, we present the details of each step in the trace transformation.
\begin{algorithm}[tpb]
\algsetup{linenosize=\tiny}
\small
\SetKwInOut{Input}{Input}
\SetKwInOut{Output}{Output}
\Input{$\langle C_1, C_2\rangle$, two critical sections in the sequential order;}
\Output{A type, indicating the ULCP type between $C_1$ and $C_2$}
  \uIf{$C_1.S_{rd}=\emptyset$ {\bf and} $C_1.S_{wr}=\emptyset$  {\bf or} $C_2.S_{rd}=\emptyset$  {\bf and} $C_2.S_{wr}=\emptyset$}{\label{alg:alg1:1}
    \Return NULL\_LOCK\;
  }
  \uElseIf{$C_1.S_{wr}=\emptyset$ {\bf and} $C_2.S_{wr}=\emptyset$}{\label{alg:alg1:2}
    \Return READ\_READ\;
  }
  \uElseIf{$C_1.S_{rd}\cap C_2.S_{wr}=\emptyset$ {\bf and} $C_1.S_{wr}\cap C_2.S_{rd}=\emptyset$ {\bf and} $C_1.S_{wr}\cap C_2.S_{wr}=\emptyset$}{\label{alg:alg1:3}
    \Return DISJOINT\_WRITE\;
  }
  \Else{
    \Return FALSE\;
  }
  \caption{ULCP Identification}
  \label{alg:alg1}
\end{algorithm}
\subsection{Building ULCP-free Topology}
\label{ULCP:trans:identification}
Prior to building ULCP-free topology, we need to identify ULCPs. We use shadow memory \cite{shadowmemory} to store the state information about critical section. Shadow memory state refers to the information about each critical section \emph{C} of the running program, which mainly consists of two sets:
\begin{itemize}
  \item $C.S_{rd}$. a set of all shared reads in the critical section $C$.
  \item $C.S_{wr}$. a set of all shared writes in the critical section $C$.
\end{itemize}

We identify ULCPs in different categories. As shown in Algorithm~\ref{alg:alg1}, null-lock, read-read, and disjoint-write can be easily identified by intersecting the read-write sets of critical sections as line~\ref{alg:alg1:1}, \ref{alg:alg1:2}, \ref{alg:alg1:3} indicate. But both benign ULCPs and true lock contention pairs (TLCPs) involve the conflicting access. In this case, Algorithm~\ref{alg:alg1} does not work. To further distinguish the false conflict of benign ULCPs from the real conflict of TLCPs, we extend the reversed replay execution~\cite{Narayanasamy2007} for the distinction between benign ULCPs and TLCPs by additionally replaying the execution trace with a reversed order of two critical section for a given ULCP. If the two replays produce the same result, then this ULCP can be classified as a benign pattern.

\begin{figure}[t]
\centering
\includegraphics[scale=0.5]{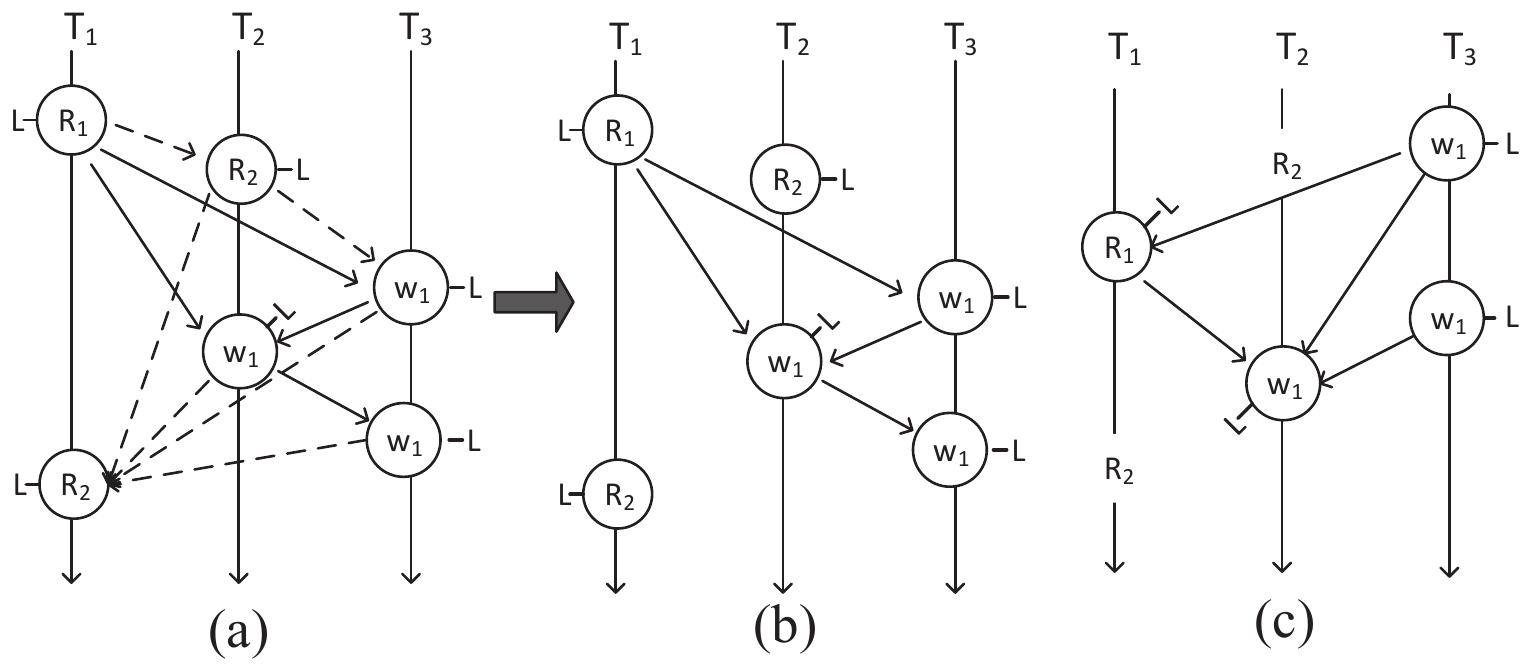}
\caption{The causal dependencies for an example. $\bigcirc$ represents the critical section, while \textsf{L} attached to $\bigcirc$ means this critical section is synchronized by lock \textsf{L}. {$\sf R_1$} indicates a read operation on shared data \textsf{1} and the dotted arrow shows a ULCP.}
\label{fig:topology}
\end{figure}

In the original topology, we know the timing relationship with respect to all critical sections in the original execution. For a certain critical section $CS$, in order to search another $CS'$ in other threads, which comprises the TLCP with $CS$, we define the operations:
\begin{itemize}
  \item \emph{\textbf{Sequential searching}} refers to searching such $CS'$ in a given thread in the order from the timing index of $CS$ to largest timing index of that thread.
  \item If we find such a $CS'$ in a given thread, it is called \emph{\textbf{matched}}.
\end{itemize}

Afterwards, we define the first rule to facilitate the building of ULCP-free topology from an original ULCP topology.
\begin{mydef}
\label{rule:1}
A causal edge is established only when the \emph{current} critical section and its first matched critical section in every other thread constitute a TLCP during the sequential searching.
\end{mydef}

Figure~\ref{fig:topology}(a) depicts an example of the building process of the ULCP free causal-order topology. To begin with, we denote the critical section {$\sf R_1$} in thread {$\sf T_1$} as the \emph{current} critical section. Then it is matched with {$\sf R_2$} in {$\sf T_2$}. $\sf R_1$ and $\sf R_2$ consist of a Read-Read ULCP. We use the dotted arrow to denote the non-causal edge relation between them. $\sf R_1$ in $\sf T_1$ is successively matched with $\sf W_1$ in $\sf T_2$, in which case there establishes a causal edge between them due to the TLCP relation, denoted as the solid arrow. When the first causal edge with $\sf W_1$ in $\sf T_2$ for $\sf T_2$ is established, $\sf R_1$ in $\sf T_1$ starts to do the similar traverse in $\sf T_2$, establishing another causal edge with the first $\sf W_1$ in $\sf T_3$. After the first round of causal edge building, $\sf R_2$ in $\sf T_2$, subsequent to $\sf R_1$ in $\sf T_1$, becomes the new \emph{current} critical section, and repeats the previous procedure.

Figure~\ref{fig:topology}(b) illustrates the ULCP-free topology built according to Rule~\ref{rule:1}. Following the program semantics of ULCP-free topology in Figure~\ref{fig:topology}(b), we may get the program execution as shown in Figure~\ref{fig:topology}(c) which affects the performance fidelity for the multiple replays (detailed discussion about this will be presented in Section~\ref{sec:perffidelity}). In order to observe the \emph{stable} performance impact of ULCPs, we then put forward Rule~\ref{rule:2}.
\begin{mydef}
\label{rule:2}
All causal-edge nodes protected by the same lock in the ULCP free topology are guaranteed with the same partial order as the original topology.
\end{mydef}

In the original topology, the partial order of the nodes $R_1$ in $T_1$, $W_1$ in $T_2$ and two $W_1$ in $T_3$ in Figure~\ref{fig:topology}(a) is \{$R_1(T_1)\prec$$W^{1st}_1(T_3)$$\prec W_1(T_2)\prec$$W^{2nd}_1(T_3)$\}.
According to Rule~\ref{rule:2}, the nodes $R_1$ in $T_1$, $W_1$ in $T_2$ and two $W_1$ in $T_3$ of ULCP-free topology in Figure~\ref{fig:topology}(b) should be restricted to the same partial order with the original topology as \{$R_1(T_1)\prec$$W^{1st}_1(T_3)$$\prec W_1(T_2)\prec$$W^{2nd}_1(T_3)$\}.

In summary, we apply RULE 1 and 2 to build the ULCP-free topology, which will be refined by RULE 3 and 4.

\subsection{Re-establishing the Program Structure of the ULCP-free Topology}
\begin{figure}[tbp] 
\centering
\includegraphics[scale=0.8]{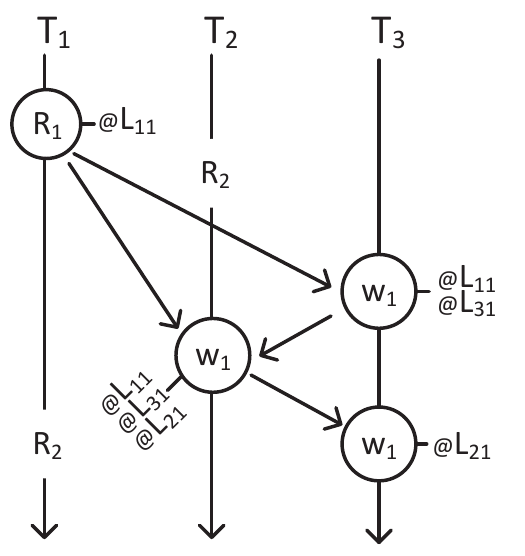}
\caption{The re-synchronization of the ULCP free causal dependencies. \textsf{@L} indicates auxiliary locks.}
\label{fig:topologyfix}
\end{figure}

We eliminate the false inter-thread dependencies caused by different categories of ULCPs. First, in absence of conflict with any critical section, \textsc{PerfPlay} removes lock/unlock events of all \texttt{null-locks} and all standalone nodes in the topology, such as $\sf R_2$ in $\sf T_1$ and $\sf R_2$ in $\sf T_2$ as shown in Figure~\ref{fig:topologyfix}(a). Second, to ensure true inter-thread dependencies between two critical sections, we use lockset~\cite{Eraser} to protect the critical sections in the topology. Lockset is a software component comprising multiple locks, which is generally used as a fine-grained lock synchronization. Consequently, \textsc{PerfPlay} uses many distinct auxiliary synchronization locks instead of the original locks to reconstruct the ULCP-free causal dependencies. Note, all these auxiliary synchronization locks provided by \textsc{PerfPlay} are written with a prefix \textsf{@L} for the discrimination from the original one.

Now, the question is on how to assign these ad-hoc locks onto each node in the ULCP-free topology while ensuring the program correctness. We perform the re-synchronization procedure as RULE~\ref{rule:3} describes.
\begin{mydef}
\label{rule:3}
Each node with the outdegree in the topology will be given a new auxiliary lock. While each node with the indegree should be synchronized by the given lock of its source node.
\end{mydef}
Figure~\ref{fig:topologyfix} shows the outcome of the example in Figure~\ref{fig:topology} according to RULE~\ref{rule:3}. According to RULE~\ref{rule:3}, the nodes with outdegrees, namely $\sf R_1$ in $\sf T_1$, $\sf W_1$ in $\sf T_2$ and $\sf W_1$ in $\sf T_3$, are given with new auxiliary $\sf @L_{11}$, $\sf @L_{21}$ and $\sf @L_{31}$, respectively. While the node with the example of $\sf W_1$ in $\sf T_3$ has the indegree from the given source node $\sf R_1$ in $\sf T_1$, thus it needs to be synchronized with the additional lock of the source node $\sf R_1$ in $\sf T_1$, i.e., $\sf @L_{11}$. Ultimately, $\sf W_1$ in thread $\sf T_3$ has the lock-set $\sf LS$=\{$\sf @L_{11},@L_{31}$\}. Each critical section will maintain a lock-set. Therefore a new mutex relationship can be described as follow:
\begin{mydef}
\label{rule:4}
Two critical sections are mutually-exclusive if the intersection of their lockset $LS$ is empty-set.
\end{mydef}

\begin{mydef1}[\textbf{Correctness}]
The transformed ULCP free trace is performed with a guarantee of either the program correctness or reporting the data races.
\end{mydef1}
\emph{Proof}. We reduce program correctness into the trace correctness due to the trace-based transformation. Consider such a typical model for the execution trace: two threads have the execution sequences \{$SG_1, A, SG_2$\} and \{$SG'_1, B, SG'_2, C, SG'_3$\}, respectively. $A$, $B$ and $C$ are the critical sections protected by the same lock, and $SG$ denotes the program segment. It is assumed that $A$ precedes $B$ in one observation, with $\langle A, B\rangle$ consisting of a ULCP and $\langle A, C\rangle$ being TLCP. According to RULE~1, $C$ is the first TLCP for $A$, a causal edge between $A$ and $C$ should be established. Therefore $SG_1$ and $SG'_3$ maintain the original semantics based on RULE~2. Only difference of ULCP-free trace from the original one lies in the parallelism between $SG_1$ and $SG'_2$ due to RULE~3.
\begin{itemize}
  \item If the segments $SG_1$ and $SG'_2$ involve the conflict free memory accesses, the ULCP free trace will be performed with the same result as the original one. Therefore, the correctness of the ULCP free trace is guaranteed in the sense that it produces the same program semantics as the original one.
  \item Otherwise, our transformation possibly produces diverse results due to the problematic interleavings of shared accesses, i.e., interleaving-sensitive data races \cite{CP:relation, acculock} between $\langle SG_1,B\rangle$, $\langle SG_1,SG'_2\rangle$, or $\langle A,SG'_2\rangle$. This case may present the correct program semantics of other executions, but it produces the same value of data races as ULCP performance problem. It further enables \textsc{PerfPlay} to help developers understand the correctness of the original trace. \qed
\end{itemize}

\begin{figure}[tbp] 
\centering
\includegraphics[scale=0.9]{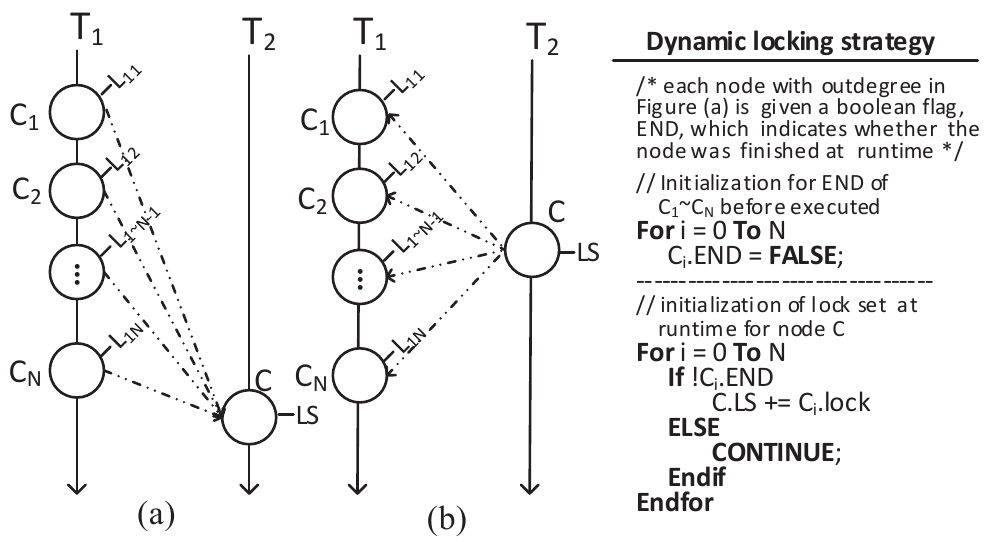}
\caption{Dynamic locking strategy}
\label{fig:dynamiclocking}
\end{figure}

One implementation detail is worthy of being further discussed. After applying RULE~\ref{rule:3}, a node in the topology may suffer from the overhead of maintaining the large-scale locksets. For instance, the lockset of the critical section $\sf C$ in Figure~\ref{fig:dynamiclocking}(a) is $\sf \bigcup_{i=0}^NL_{1i}$. To reduce runtime overhead of the large lock-sets, we propose a dynamic locking strategy, as shown in Figure~\ref{fig:dynamiclocking}, the main idea of which is that the synchronization of the targeted node $\sf C$ depends on the runtime state (i.e., \textsf{END}) of each source node $\sf C_1, \cdots, C_N$. For instance, if $\sf C_1.END=TRUE$, it means that the critical section $\sf C_1$ is already finished. If the node $\sf C_1$ is finished before the execution of the node $\sf C$ at runtime, the lockset $\sf LS$ of the node $\sf C$ can exclude the lock of one of its source nodes $\sf C_1$, i.e., $\sf L_{11}$. Based on the dynamic locking strategy, \textsc{PerfPlay} saves much overhead for the maintenance of the locksets and is able to deal with any thread interleaving as shown in Figure~\ref{fig:dynamiclocking}(b).

\section{ULCP Performance Debugging}
\label{sec:replay}
After the phase of ULCP transformation, we obtain a set of ULCPs. For an effective debugging framework, we still face one major problem. There may be many ULCPs, and some of them are even from the same code-site. An effective debugging tool should point out the succinct code-site for distinctive ULCPs, and also locate the most performance critical ULCP for programmers. Thus, we propose ULCP fusion and performance accumulation based on their code regions in the source code level (Section~\ref{sec:perf:impro}), and point out the most performance critical ULCP to programmers (Section~\ref{sec:priortization}).

\begin{figure}[t] %
\centering
\includegraphics[scale=0.9]{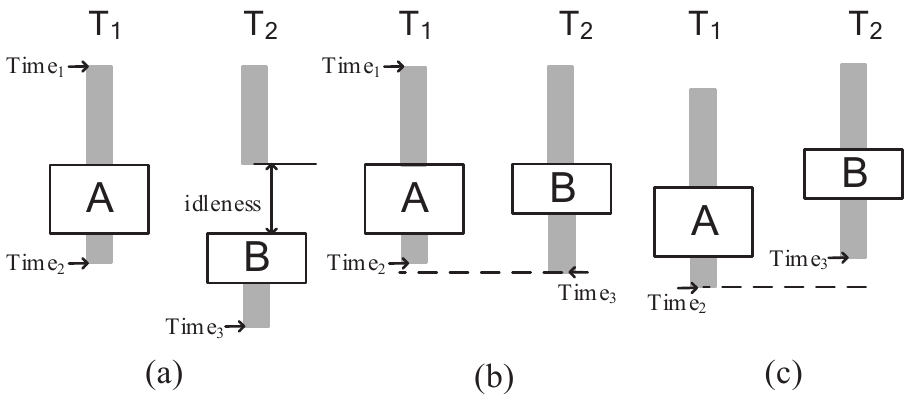}
\caption{Two different performance measurements}
\label{fig:perfmeasure}
\end{figure}
\subsection{ULCP Fusions}
\label{sec:perf:impro}
We model the potential runtime overhead of a ULCP. Figure~\ref{fig:perfmeasure} illustrates a detailed diagrammatic representation of the performance metrics, where $A$ and $B$ constitute a ULCP. We label the start point of precursor segment of the first critical section $A$ using $Time_1$; the end point of successor segment of $A$ using $Time_2$; the end point of successor segment of the second critical section $B$ using $Time_3$. When the ULCP free trace is executed, the replayed program may perform the traces in two possible ways, as shown in Figure~\ref{fig:perfmeasure}(b) and Figure~\ref{fig:perfmeasure}(c). We consider both cases: for case (b), the improved performance of ULCP is $\Delta Time_3$$-$$\Delta Time_1$; for case (c), the result is $\Delta Time_2$$-$$\Delta Time_1$. Consequently, we define the performance improvement of each ULCP as follows.
\begin{equation} \label{equ:ilcp}
\setlength{\abovedisplayskip}{10pt} 
\setlength{\belowdisplayskip}{10pt} %
\Delta T_{ULCP}=\Delta MAX\{Time_{2},Time_3\}-\Delta Time_{1}
\end{equation}
where $Time_{label}$ indicates the current timestamp of application when the program is executed at the location of $label$, and $MAX$ is denoted as the maximum value. $\Delta$ is denoted as an operation that calculates D-value (difference value) before and after optimization.

After the process of Algorithm~\ref{alg:alg1}, \textsc{PerfPlay} collects a large number of ULCPs, denoted as \{$ULCP_1,ULCP_2,\cdots,ULCP_n$\}, each consisting of two critical sections $\langle C_1, C_2\rangle$. To facilitate the description, we define the operator $\cdot$ to obtain the attribute or component of a ULCP, such as $ULCP_1.C_1$. However, some ULCPs are possibly caused by the same code region (CR). Thus, we propose ULCP fusion to merge two ULCPs into the unique ULCP per code region in the source code level. Then, we can report to programmers the accumulated performance impact of ULCPs at the CR level. Particularly, we accumulate up the performance improvement of ULCPs generated by the same code regions according to Algorithm~\ref{alg:alg2}. In Algorithm~\ref{alg:alg2}, $\langle CR_1, CR_2\rangle$ is denoted as the code regions incurring two critical sections $\langle C_1, C_2\rangle$ of a ULCP. The binary operator $\sqcup$ means whether two $CR$s involve the shared region of the code; while $\sqcap$ indicates the conflated code region of two $CR$s. Through Algorithm~\ref{alg:alg2}, the final state of ULCP group is that any two ULCPs can not be fused further.
\begin{algorithm}[tpb]
\algsetup{linenosize=\tiny}
\small
\SetKwInOut{Input}{Input}
\SetKwInOut{Output}{Output}
\Input{$\sf \langle ULCP_1, ULCP_2\rangle$, two standalone ULCPs;}
\Output{$\sf ULCP_{new}$, a new synthetic ULCP; \\ {\sf NULL}, two standalone ULCPs that can not be merged}
  \tcc{Handle the same code regions or nested locks}
  \uIf{$\sf ULCP_1.CR_1 \sqcap ULCP_2.CR_1\neq\emptyset$ {\bf and} $\sf ULCP_1.CR_2 \sqcap ULCP_2.CR_2 \neq\emptyset$}{\label{alg:alg2:1}
     $\sf ULCP_{new}.CR_1 \leftarrow ULCP_1.CR_1 \sqcup ULCP_2.CR_1$\;
     $\sf ULCP_{new}.CR_2 \leftarrow ULCP_1.CR_2 \sqcup ULCP_2.CR_2$\;
     $\Delta T_{\sf ULCP_{new}}\leftarrow\Delta T_{\sf ULCP_1}$+$\Delta T_{\sf ULCP_2}$\;
  }
  \uElseIf{$\sf ULCP_1.CR_1 \sqcap ULCP_2.CR_2\neq\emptyset$ {\bf and} $\sf ULCP_1.CR_2 \sqcap ULCP_2.CR_1$ $\neq\emptyset$}{\label{alg:alg2:2}
     $\sf ULCP_{new}.CR_1 \leftarrow ULCP_1.CR_1 \sqcup ULCP_2.CR_2$\;
     $\sf ULCP_{new}.CR_2 \leftarrow ULCP_1.CR_2 \sqcup ULCP_2.CR_1$\;
     $\Delta T_{\sf ULCP_{new}}\leftarrow\Delta T_{\sf ULCP_1}$+$\Delta T_{\sf ULCP_2}$\;
  }
  \Else{
     $\sf ULCP_{new} \leftarrow NULL$\;
  }
  \caption{ULCP Fusion and Performance Accumulation}
  \label{alg:alg2}
\end{algorithm}
\subsection{ULCP Recommendations}
\label{sec:priortization}
After ULCP fusion and performance accumulation by Algorithm~\ref{alg:alg2}, we obtain a group of unique ULCPs, denoted as \{$ULCP_1,ULCP_2,$ $\cdots,ULCP_m$\}, and its corresponding performance improvement \{$\Delta T_{ULCP_1},$ $\Delta T_{ULCP_2}, \cdots, \Delta T_{ULCP_m}$\}. For the effectiveness of a debugging tool, it is desirable to prioritize the most beneficial ULCPs to programmers, since there may be many ULCPs in the program.
We denote $P$ as the relatively optimizable value of a ULCP among the total ULCP group:
\begin{equation}
\setlength{\abovedisplayskip}{10pt} 
\setlength{\belowdisplayskip}{10pt} %
 P = \frac{\Delta T_{ULCP}}{\sum^m_{j=1}\Delta T_{ULCP_j}}
\end{equation}
which represents the relative optimization opportunity of a corresponding ULCP. Each ULCP in \{$ULCP_1,ULCP_2,\cdots,ULCP_m$\} has its own $P$, and $\sum_{i=1}^m ULCP_i.P=1$. To further ascertain the most beneficial ULCPs, we resort \{$ULCP_1,ULCP_2,\cdots,ULCP_m$\} by $P$ in a descending order, i.e., $\forall i>j, ULCP_i.P<ULCP_j.P$. Then we can pinpoint the most performance critical code regions from that order list. Thus, our tool can recommend the most performance critical ULCP as $ULCP_1$.

\section{Implementation Issues}
\label{sec:perfplay}
We implement \textsc{PerfPlay} for the parallel replay based on Pin \cite{PIN}, an underlying framework that enables programmers to perform the program analysis at runtime without source codes. In particular, we remove and insert the auxiliary locks in the trace level, instead of modifying the compiler or binary. Modifying trace with the lock mechanisms can provide an easy implementation for that objective, which has the same effect to provide useful debugging hints for ULCPs as modifying the binary or compiler.
This section focuses on two implementation details: i) what information should be recorded for the performance analysis using replay technique in the recording phase; ii) how to perform the faithful replay for each run upon the given trace so that the performance impact of examined problems can be evaluated precisely in the replay phase.
\begin{figure}[tbhp] 
\centering
\includegraphics[scale=0.9]{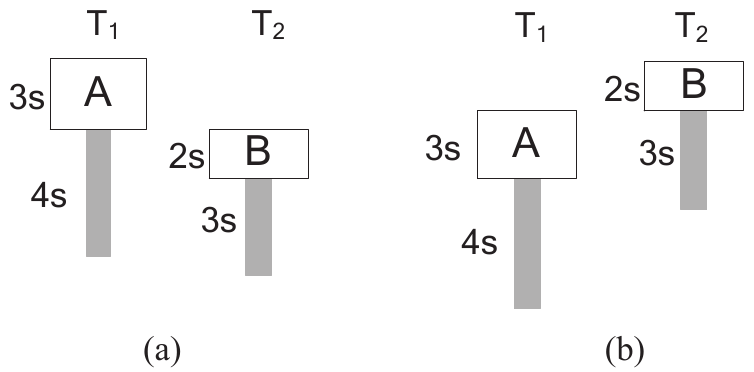}
\caption{An example of the lock mutual exclusion for the performance fluctuation with different sequences, where the digits indicate the time cost of the program segment. (a) If \textsf{A} precedes \textsf{B}, the program costs $8$s; (b) If \textsf{B} precedes \textsf{A}, the program costs $9$s.}
\label{fig:perffidelity}
\end{figure}
\subsection{What and How to Record}
\label{sec:recording}
To evaluate the performance impact of ULCP, we should record all information of ULCP to perform its performance. Thus, it is necessary to record all instructions and memory accesses between lock and unlock operations. 

For other events, the recording strategy of \textsc{PerfPlay} is quite flexible, ranging from complete recording to selective recording. Complete recording refers to the recording of the complete execution of a program, with complete performance analysis; while selective recording \cite{pinplay} is proposed to reduce the recording overhead and accelerate the replay speed, which is generally used for the local performance analysis. The basic idea of selective recording is to only record the changes of the states and values of memory (including registers and RAM) before and after running a specific code range, such as a system call, library calls and spin-loop. During the process of replay, the specific code range is bypassed, and the relevant states and values of memory are restored instead of being executed again. Thus, \textsc{PerfPlay} chooses selective recording whenever appropriate. Specifically, for non-mutual exclusive semaphores, \textsc{PerfPlay} only ensures the correctness of the partial order in the sense that it is the same as the original ordering. On the other hand, with the purpose of further facilitating the repeated debugging of ULCPs, \textsc{PerfPlay} also supports the checkpoints, which is useful for programmers to focus on a smaller code region.

\subsection{Performance Fidelity}
\label{sec:perffidelity}
There has been significant amount of work on building record/replay systems~\cite{Narayanasamy2007, MutableReplay, pinplay, BugNet} for understanding the correctness of bugs in programs, but not much effort has gone into leveraging them to study performance issues. Based upon a given trace, the determined information contains: the path branches each thread performs, synchronization operations, and the instructions or events performed by each thread. Therefore, suppose we perform the same trace twice, performance fluctuation of the program largely depends on the lock synchronization interleaving. As shown in Figure~\ref{fig:perffidelity}, if two critical sections coequally contend for the lock resource, the program may perform different performance due to the potential different time cost of subsequent program segments.

To enable performance analysis using replay technique (\emph{abbr}. performance replay) for the parallel execution, we propose an enforced locking serialization constraint (ELSC) which enforces the total order of the dynamic lock synchronizations for the replayed trace according to the schedule order of these locks at runtime. That is, ELSC schedules the same lock order as the scheduled order of these locks when the program runs at runtime. As shown in Figure~\ref{fig:perffidelity}, if the program runs as Figure~\ref{fig:perffidelity}(a) shows when the trace is being recorded, {ELSC sets down this order of $A\rightarrow B$ in the recording phase} and then enforces ALL subsequent replays for this trace with \emph{hard} ordering of \textsf{A} happening before \textsf{B} in the replay phase. ELSC ensures the performance fidelity of replay execution for the multiple replays based upon the same given trace.

\begin{mydef1}[\textbf{Performance Fidelity}]
ELSC guarantees the performance fidelity of the parallel replay for the same given trace.
\end{mydef1}
\emph{Proof}. Consider such a generic model of lock synchronization: two threads have the execution sequences $\{SG_1, A, SG_2\}$ and $\{SG'_1, B, SG'_2\}$, respectively. $A$ and $B$ are the critical section protected by the same lock. $SG$ denotes the program segment. Supposing $A$ precedes $B$ in one observation $\tau_0$, and $\{SG_1, A, SG_2, SG'_1,$ $B, SG'_2\}$ cost the execution time $\{t_1, t_A, t_2,$ $t'_1, t_B, t'_2\}$ respectively. 
There are three cases for the subsequent replays:
\begin{itemize}
  \item If $t_1>t'_1$, the critical section $B$ will first require the lock and this result is in violation with the observation $\tau_0$.
  \item If $t_1<t'_1$, the critical section $B$ has to wait for the lock release of $A$ in each run, yielding the same execution as $\tau_0$.
  \item If $t_1=t'_1$, $A$ and $B$ will contend for the acquisition of lock. Getting the preferential lock depends on the system scheduling. ELSC constrains this case with the preferential lock for $A$ based on $\tau_0$, thereby consequently triggering the execution $\tau_0$.
\end{itemize}

In most cases, it is common that multiple threads contend \emph{fairly} (i.e., $t_1=t'_1$) for the same lock release of another thread. To guarantee the performance consistency with $\tau_0$, ELSC assigns the sequences of lock acquisition for these threads based on the historical trace $\tau_0$.
A multi-threaded program is usually consisted of many such generic models such that the performance fidelity of the execution trace is guaranteed. \qed

\begin{figure}[tpb] 
\centering
\includegraphics[scale=1.1]{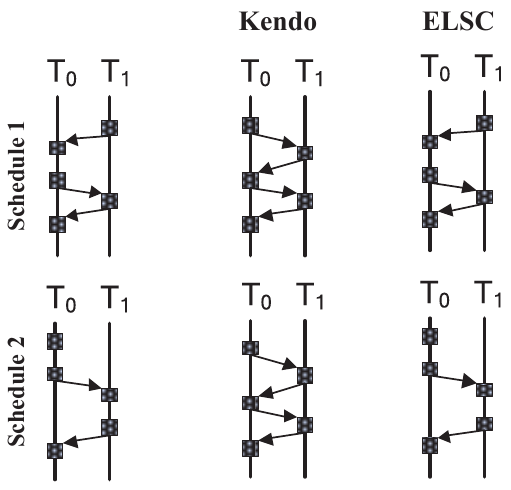}
\caption{The difference of ELSC and Kendo for the synchronization order enforcement. Both \textsf{Schedule 1} and \textsf{Schedule 2} of program are performed under the same input.}
\label{fig:RSC}
\end{figure}

To ensure the determinism of program execution, both PinPlay~\cite{pinplay} and CoreDet~\cite{CoreDet} enforce the same total order of all shared memory access points under the same input. They only relax the parallelism of private memory accesses for each thread, thus slowing down the program execution significantly by 2X-20X \cite{pinplay,CoreDet}.  We also note one related work named Kendo~\cite{Kendo}, which enforces the deterministic order of all lock acquisitions under the same input. Figure~\ref{fig:RSC} illustrates the difference between ELSC and Kendo. In Figure~\ref{fig:RSC}, we can see Kendo always enforces the fixed synchronization order for the same input regardless of the runtime scheduling of program. That is, Kendo will enforce the same lock order for both {\sf Schedule 1} and {\sf Schedule 2}, because \textsf{Schedule 1} and \textsf{Schedule 2} are performed under the same input. However, this strategy generally extends the execution time of programs~\cite{Kendo}, thereby increasing the uncertainty for performance analysis. For instance, if the program performs as \textsf{Schedule 1}, at this moment, Kendo has to defer the execution of the first critical section in $\sf T_1$ until the end of the first critical section in $\sf T_0$, and \textsf{Schedule 2} also has the time extension with the same principle. Unlike the \emph{input-driven} feature of Kendo, ELSC performs the \emph{schedule-driven} strategy for the synchronization order, i.e., ELSC only provides the determinism for the same schedule. For the instance in Figure~\ref{fig:RSC}, if \textsf{Schedule 1} is scheduled by the program when the trace is being recorded, ELSC will enforce the same synchronization order as the \textsf{Schedule 1} when replaying this trace. While the trace performed with \textsf{Schedule 2} will be enforced with a different lock order by ELSC. The schedule-driven characteristic of ELSC strictly adheres to the original schedule of real program execution, and eliminates the nondeterminism of the enforced lock waiting time in Kendo, thus producing the faithful performance.
\section{Evaluation}
\label{sec:evaluation}
\subsection{Experimental Setup}
\textbf{System configuration:} All experiments are performed on a machine with two Intel quadcore Xeon E5310 1.60Ghz processors, 8GB memory, one 250GB SATA hard disk, and 1Gbit Ethernet interface. The operating system was CentOS 5.6 (X86\_64) with Linux kernel 3.0.0-12.

\textbf{Benchmark test configuration:}
We evaluate \textsc{PerfPlay} using two servers (\emph{openldap} \cite{OpenLDAP} and \emph{mysql} \cite{mysql}), three desktop applications (\emph{pbzip2} \cite{pbzip2}, \emph{transmissionBT} \cite{transmissionbt} and \emph{handBrake} \cite{handbrake}) and PARSEC benchmarks \cite{parsec}. 

1) \emph{openldap}: a lightweight directory access protocol server. In our test, we use the default thread pool mode for \emph{openldap} server, and use the professional tool DirectoryMark by MindCraft\footnote{http://www.mindcraft.com/directorymark/} to benchmark it with the option of searching $2000$ entries.

2) \emph{mysql}: an open source database system which is widely-used in the world. We use the test tool mysqlslap released in \emph{mysql} software package to test \emph{mysql} with $1000$ queries, and $2$ iteration.

3) \emph{pbzip2}: a parallel implementation of the bzip2 compressor. We test it by compressing a $256$M file with two processors.

4) \emph{transmissionBT}: a BitTorrent client. We only test its download function by downloading a local 300M file.

5) \emph{handBrake}: a video transcoder. We test it by conversing a $256$M  DVD format file into MP$4$ format with H.$264$ codec, $30$ FPS.

6) \emph{PARSEC Benchmarks}: a benchmark suite with $12$ multi-threaded programs. We test all PARSEC benchmarks (except \emph{freqmine}) with \emph{simlarge} input. \textsc{PerfPlay} is implemented currently based on \textsf{pthread} library. As \emph{freqmine} benchmark is an \textsf{openMP} program, \textsc{PerfPlay} can not identify its synchronization. 

\textbf{Methodology:} To demonstrate the performance fidelity of \textsc{PerfPlay}, we perform the replay execution with the following four schemes:

1. Memory-based schedule (MEM-S)~\cite{pinplay}, which enforces a deterministic execution sequence of all shared memory accesses.

2. Synchronization-based schedule (SYNC-S)~\cite{Kendo} that enforces the total order of locks for the same input.

3. ELSC-based schedule (ELSC-S), which enforces the total order of locks for the same schedule.

4. Parallel replay for the original execution without any enforcement strategy for the events (ORIG-S).

We focus on the key part of dynamic executions in the trace replay. In our implementation, we have decoupled the replayed execution time of program from the other time-consuming manipulations, such as the loading of trace from disk into memory, and the trace format transformation from the string-style into the instruction-style.
\begin{figure}[tbp] 
\centering
\includegraphics[scale=0.35]{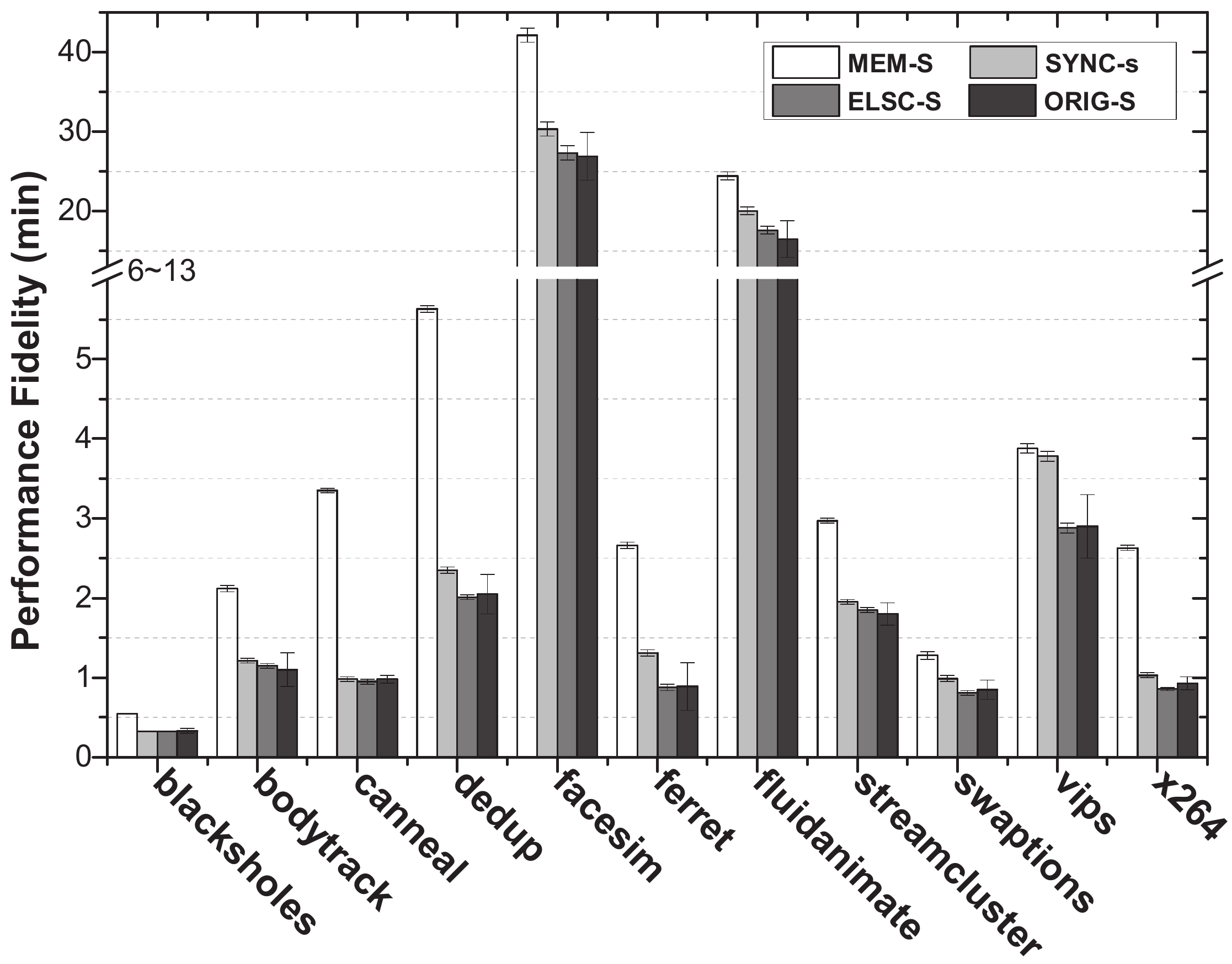}
\caption{Performance fidelity comparison between different execution schemes for the replay}
\label{fig:performancefidelity}
\end{figure}
\begin{figure}
\centering
\includegraphics[scale=0.35]{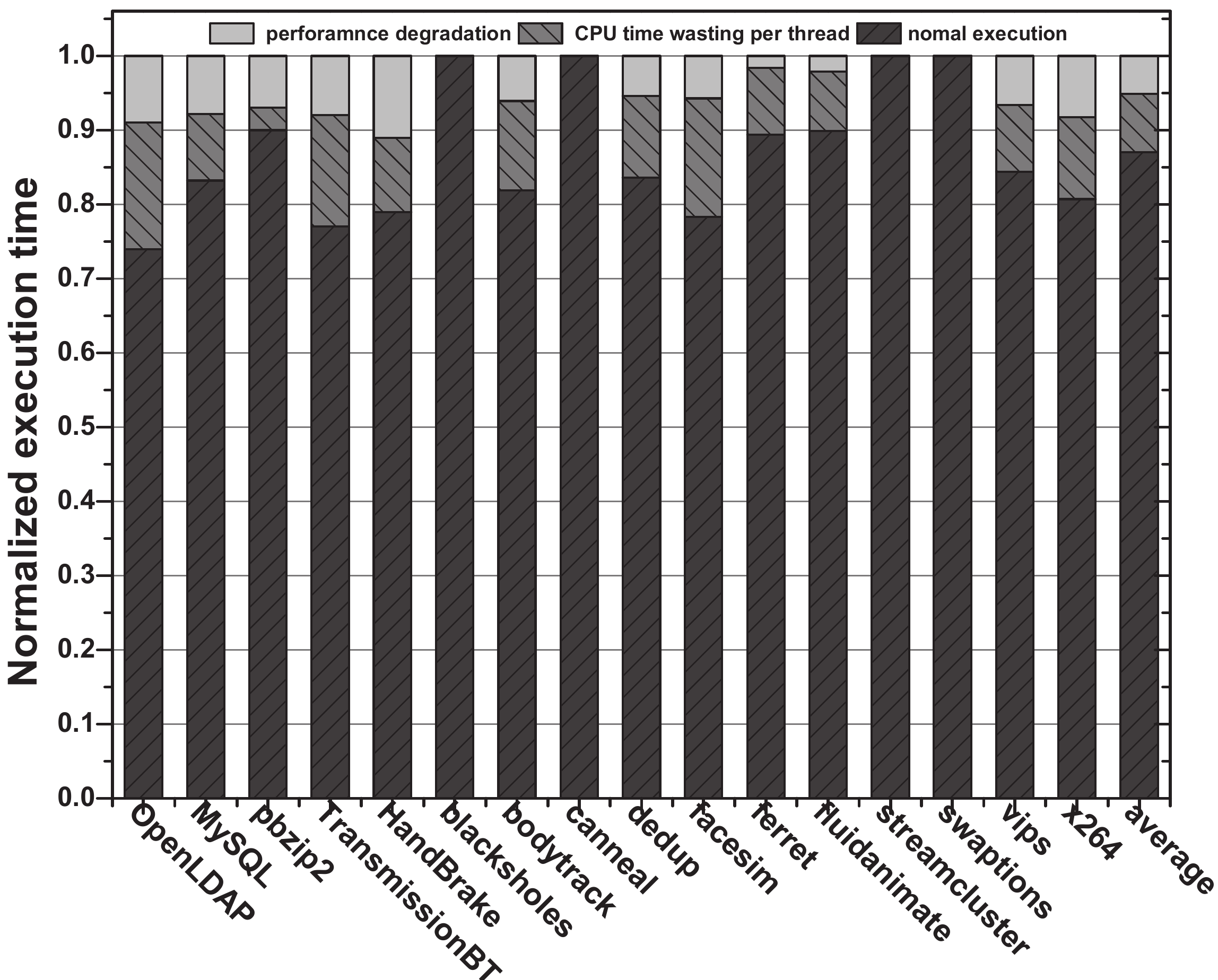}
\caption{The normalized execution time through replaying the traces with and without ULCPs}
\label{fig:overlap}
\end{figure}
\subsection{Performance Fidelity of \textsc{PerfPlay}}
\label{sec:perf:fidelity}
To evaluate performance fidelity, two aspects are needed to be assessed, including performance stability and precision. Stability represents whether \textsc{PerfPlay} shows the same performance across the multiple replays with the same trace. The precision means whether \textsc{PerfPlay} strictly adheres to the original execution. If our debugging framework has a high precision, we can determine that the performance improvement of ULCP-free replayed execution comes entirely from the optimization of ULCPs.

We record all PARSEC benchmarks with \emph{simlarge} input, and we replay the trace of each application ten times using different replay schemes (i.e., MEM-S, SYNC-S, ELSC-S, and ORIG-S). Figure~\ref{fig:performancefidelity} shows the final replayed execution time using these schemes. {From the small error bars, we can see that MEM-S, SYNC-S, and ELSC-S all enforce the deterministic program execution for the multiple times}, thus providing the \emph{stable} performance analysis. Nevertheless, ORIG-S shows the indeterminate {(large error bars)} program execution due to the inter-thread lock interleaving. Except the nature of enforcement scheme itself, both MEM-S and SYNC-S manifest themselves with the additional performance introduction compared with ORIG-S. While ELSC-S eliminates the waiting time of SYNC-S for lock acquisition by only enforcing the synchronization order based on the scheduled synchronization order for the same schedule. As a result, we can see that ELSC-S almost produces the same program performance with ORIG-S. This yields the conclusion that \textsc{PerfPlay} with ELSC scheme strictly schedules the replay execution as the original scheduled execution without any additional performance introduction, thus providing the \emph{precise} performance analysis. {From the above-discussed results, it is revealed that only ELSC-S provides both the performance stability and performance precision, thus ensuring the performance fidelity of replay execution.}

\begin{table}[tbp]   
\caption{\# Grouped ULCP code regions and optimization opportunity of the most beneficial one}
\small
\tabcolsep=0.1cm
\setlength{\abovecaptionskip}{10pt}
\setlength{\belowcaptionskip}{-10pt}
\centering
\begin{tabular}{|c|c|c|}\hline
\textbf{Applications}& \minitab[c]{\#Grouped\\ULCPs}& \textbf{$\sf ULCP_1.P$} \\ \hline \hline
{\sf openldap}&18 &30.1\%  \\
{\sf mysql}&57 &12.5\%  \\\hline
{\sf pbzip2}&4 &59.4\%  \\
{\sf transmissionBT}&2 &53.5\% \\
{\sf handbrake}&29 &15.4\%  \\ \hline
{\sf blackscholes}&0 &0  \\
{\sf bodytrack}&5 &20.9\%  \\
{\sf facesim}&11 & 31.2\% \\
{\sf fluidanimate}&3 &26.5\%  \\
{\sf swaptions}&0 &0  \\ \hline
\end{tabular}
\label{table:region}
\end{table}
\subsection{Performance Impact Evaluation}
\label{sec:perf}
Performance impact of ULCPs in this work includes:
\begin{itemize}
  \item \textbf{Performance degradation} ($T_{pd}$): The performance improvement of program before and after the optimization;
  \item \textbf{Resource wasting} ($T_{rw}$): In our test, resource wasting mainly refers to the wasting of CPU resource, which makes the useless ULCP computation (e.g., spin-lock) on the non-critical path.
\end{itemize}
\noindent where $T_{pd}$ 
can be directly quantified by replaying ULCP trace ($T_{ut}$) and ULCP-free trace ($T_{uft}$), i.e., $T_{pd}=T_{ut}-T_{uft}$. With Equation~\ref{equ:ilcp}, $T_{rw}$ can be indirectly calculated as $\sum \Delta ULCP -T_{pd}$. To quantify them in the following experiments, we evaluate them with the metric of the normalized performance impact (i.e., ${T_{pd}}/{T_{real}}$) and CPU-time wasting per thread on average (${T_{rw}}/{N_{thread}}$), respectively.
All tests are executed with two threads. 

\textbf{Performance impact of ULCPs:} Figure~\ref{fig:overlap} illustrates the normalized performance impact and normalized CPU-time wasting of ULCPs from $5$ real world programs and PARSEC benchmarks. In our tests, \textsc{PerfPlay} produces different opportunities of performance impact for different applications. For example, \emph{blacksholes}, \emph{canneal}, \emph{streamcluster}, and \emph{swaptions} hardly obtain any performance impact due to the correct use of lock or exclusive use of lock. While for other applications, such as \emph{openldqp}, \emph{mysql}, \emph{pbzip2}, the program has a significant percent for the improvement of performance ($1.6\%$--$11\%$) and CPU time per thread ($1.1\%$--$16.7\%$) due to the ULCPs. On average, the performance of these applications can be improved by $5.1\%$ and the resource utilization per thread by $7.85\%$. Usually, a program with more ULCPs has larger performance improvement, which indicates the benefits of removing ULCPs by our performance debugging framework. One exception is that, \emph{fluidanimate} has a larger number of ULCPs than \emph{facesim}, but produces a lower speedup. That is because ULCPs in \emph{facesim} have the larger-scale critical sections.

\textbf{Performance gain from the most beneficial ULCPs:} Table~\ref{table:region} reports the number of the exploited ULCP code regions and corresponding performance gain of the most beneficial one.
Column \emph{grouped ULCPs} counts total number of the unique ULCPs after the fusion and performance accumulation of ULCPs.
Column $ULCP_1.P$ (discussed in Section~\ref{sec:priortization}) shows the relative optimization portion of the most beneficial ULCP code regions among the total ULCP group set.
From Table~\ref{table:region}, we find that different applications show different optimization opportunities. For instance, \emph{openldap} has 18 grouped ULCP code regions while its most beneficial one takes up 30.1\% of optimization gain among the total ULCP set. \emph{mysql} produces a larger number (57), but the most beneficial one exhibits only 12.5\% of performance benefit.
The performance gain of the most beneficial ULCP code region for other applications can be seen in Table~\ref{table:region}.

\begin{table}[tbp]   
\caption{Runtime overhead of locksets with/without dynamic locking strategy}
\small
\tabcolsep=0.1cm
\setlength{\abovecaptionskip}{10pt}
\setlength{\belowcaptionskip}{-10pt}
\centering
\centering
\begin{tabular}{|c|c|c|}\hline
\textbf{Applications}& \textbf{w/o DSL}& \textbf{w/ DSL} \\ \hline \hline
{\sf blackscholes}& 0& 0 \\
{\sf bodytrack}& 5.3\% & 0.5\% \\
{\sf canneal}& 0.2\%& 0.2\%\\
{\sf dedup} &4.6\% & 0.7\%\\
{\sf facesim}&7.8\% &1.2\%  \\
{\sf ferret}& 10.7\%&3.6\% \\
{\sf fluidanimate}& 14.1\%& 4.3\%  \\
{\sf streamcluster}& 2.9\%&0.6\% \\
{\sf swaptions}& 0.4\%& 0.4\% \\
{\sf vips} &7.6\% & 2.4\% \\
{\sf x264}& 5.0\%& 1.9\% \\ \hline
\end{tabular}
\label{table:dsl}
\end{table}
\subsection{Overhead Reduction via Dynamic Locking Strategy}
Lockset is introduced to transform ULCPs into the parallel pattern. However, it also introduces the significant overhead for the determination of mutex relationship by intersecting two locksets in RULE~\ref{rule:4}, especially for the lock intensive programs. To quantify lockset (LS) overhead, we replay PARSEC benchmarks with and without dynamic locking strategy (DLS), respectively. Table~\ref{table:dsl} compares runtime overhead of locksets with and without DLS. When not using DSL, lockset maintenance incurs significant ($0.2\%-14.1\%$) amount of runtime overhead. In contrast, lockset with DLS further reduces performance impact of lockset into a negligible level, only incurring $4.3\%$ overhead even for the lock intensive application \emph{fluidanimate} which makes extensive use of locks.

\begin{figure}[tpb]
\centering
 \subfloat[The performance loss with the increasing number of threads]{\includegraphics[scale=0.17]{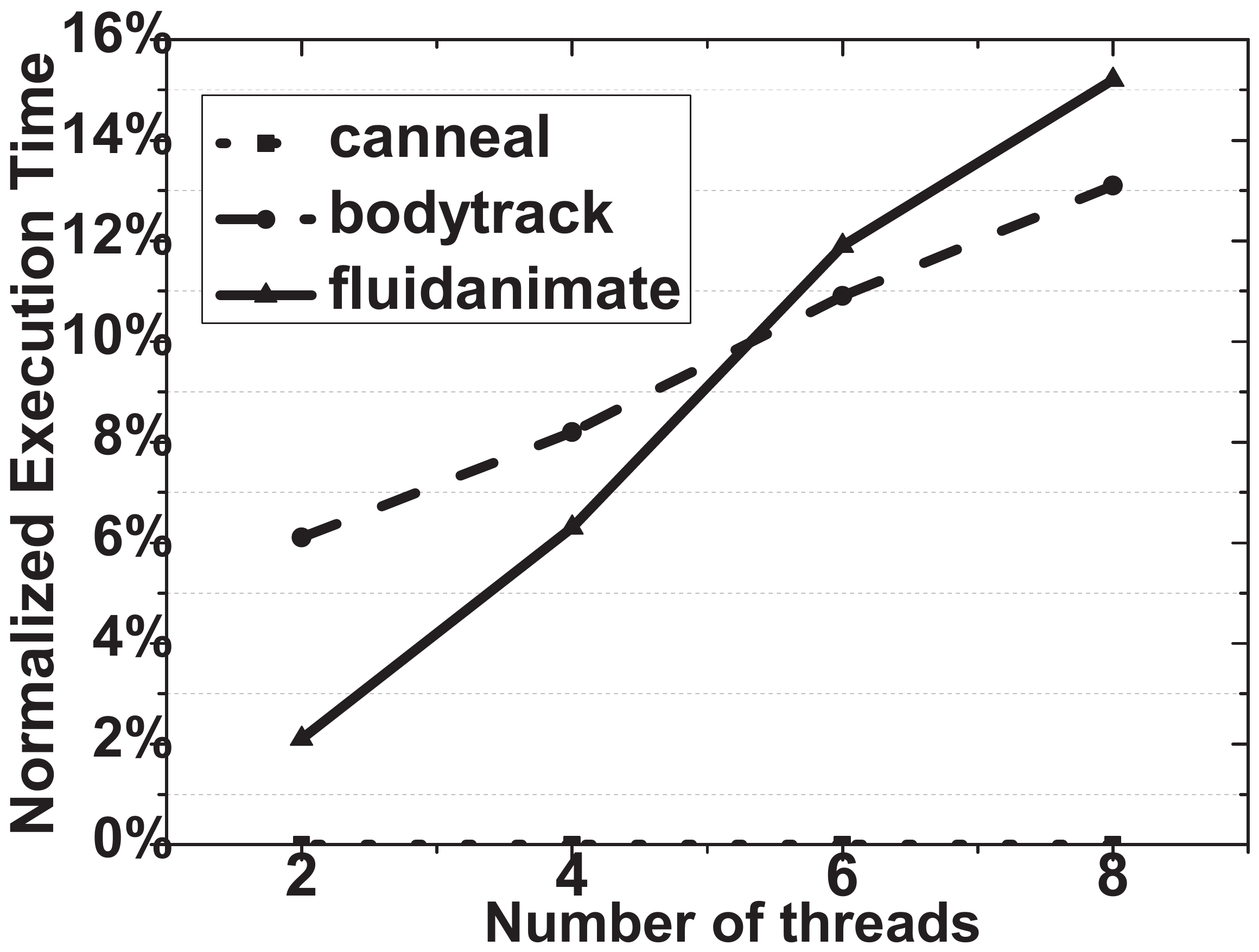}}
 \hfill
 \subfloat[The CPU wasting with the increasing number of threads]{\includegraphics[scale=0.17]{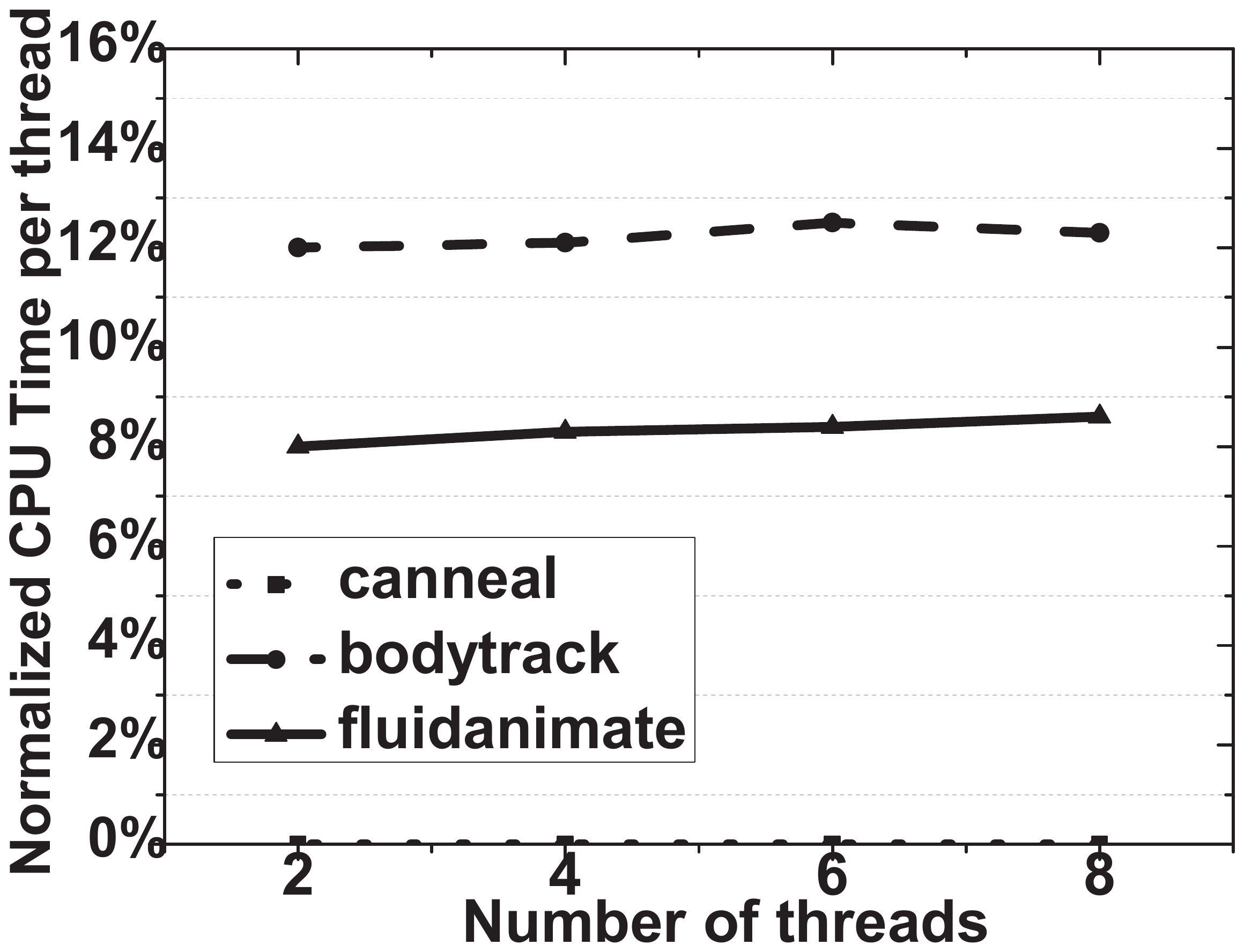}}
 \caption{ULCP impact with the increasing number of threads}
 \label{fig:perfImpact:threadnum}
\end{figure}
\begin{figure}[tpb]
\centering
 \subfloat[The performance loss with the varying input size]{\includegraphics[scale=0.16]{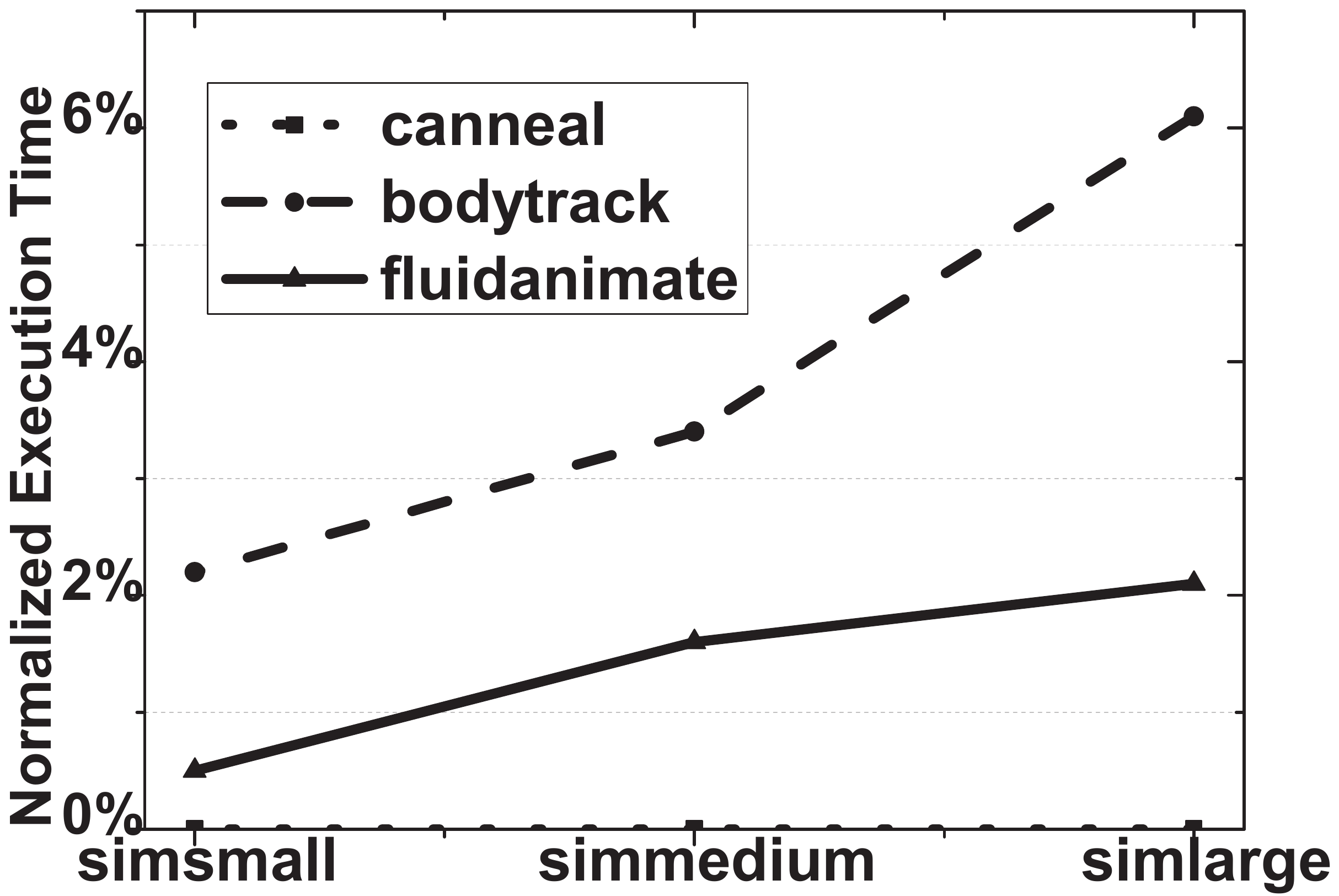}}
 \hfill
 \subfloat[The CPU wasting with the varying input size]{\includegraphics[scale=0.16]{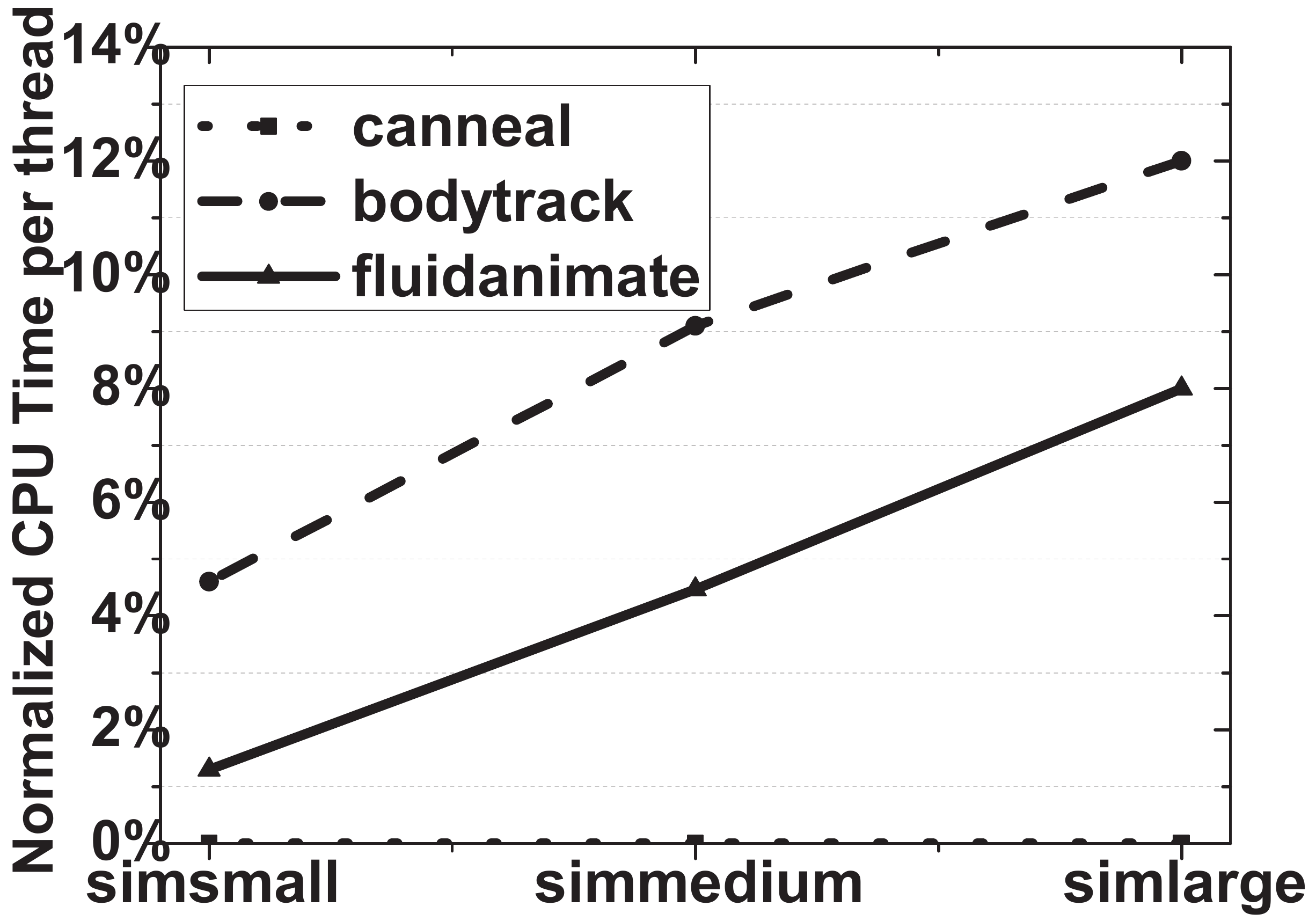}}
  \caption{ULCP impact with the varying input size}
 \label{fig:perfImpact:input}
\end{figure}
\subsection{Sensitivity Study of ULCPs}
\label{lable:sensitivity}
To evaluate the evolution of ULCP impact, we study the ULCP sensitivity to the thread number and input size. We select \emph{canneal}, \emph{bodytrack}, \emph{fluidanimate} from PARSEC benchmarks with different numbers (i.e., a few, medium, large) of ULCPs.

Figure~\ref{fig:perfImpact:threadnum} depicts the sensitivity of ULCPs to the thread number. We can find that ULCPs lead to the increasing performance loss as the number of threads increases while the resource wasting per thread stays the same.
Figure~\ref{fig:perfImpact:input} depicts the sensitivity of ULCPs to the input size. It can be observed that both performance loss and resource wasting increase as the input size increases.
The explanation for both figures is: in those applications 1) all threads reuse the same code (e.g., functions) to perform the program execution; 2) more input sizes merely mean the number of executions on some code segments is increasing.

Note that \emph{canneal} still does not show any potential opportunity with both the increasing thread number and input size. Combining the results from \emph{bodytrack} with \emph{fluidanimate}, we seems to reveal that in most cases the ULCP code-sites are not affected by the thread numbers and input sizes of these applications. ULCPs can manifest themselves in two threads, and more thread numbers may only change their performance impact.
\begin{figure}[t]
\centering
\begin{tabular}{c}
\begin{lstlisting}[escapechar=@,linewidth=8cm]
int Query_cache::try_lock(bool){
  @\textbf{mysql\_mutex\_lock}@(&structure_guard_mutex);
  while(1){
  set_timespec_nsec(waittime,(@\textbf{\textcolor[rgb]{0.00,0.00,1.00}{ulong}}@)5000000L);
  int res=mysql_cond_timedwait(
            &COND_cache_status_changed,
            &structure_gurad_mutex,&waittime);
  if(res==EITMEOUT){
     ...
     break;
  }
  }
  @\textbf{mysql\_mutex\_unlock}@(&structure_guard_mutex);
}
\end{lstlisting}
\end{tabular}
\caption{A verified ULCP problem from \textsf{mysql}}
\label{fig:official:case}
\end{figure}
\begin{figure}[t]
\centering
\begin{tabular}{c}
\begin{lstlisting}[escapechar=@,linewidth=8cm]
void *consumer(void *q){
2109:   @\textbf{pthread\_mutex\_lock}@(&mu);
2122:   if(fifo->empty&&syncGetProducerDone()==1)
2124:         @\textbf{pthread\_mutex\_unlock}@(&mu);
     }
int syncGetProducerDone(){
 533:     int ret;
 534:     @\textbf{pthread\_mutex\_lock}@(&muDone);
 535:     ret=producerDone;
 536:     @\textbf{pthread\_mutex\_unlock}@(&muDone);
 537:     return ret;
 538: }
\end{lstlisting}
\end{tabular}
\caption{A ULCP problem from \texttt{pbzip2}}
\label{fig:pbzip}
\end{figure}
\begin{figure}[htpb]
\centering
 \subfloat[Case study with the varying number of threads]{\includegraphics[scale=0.17]{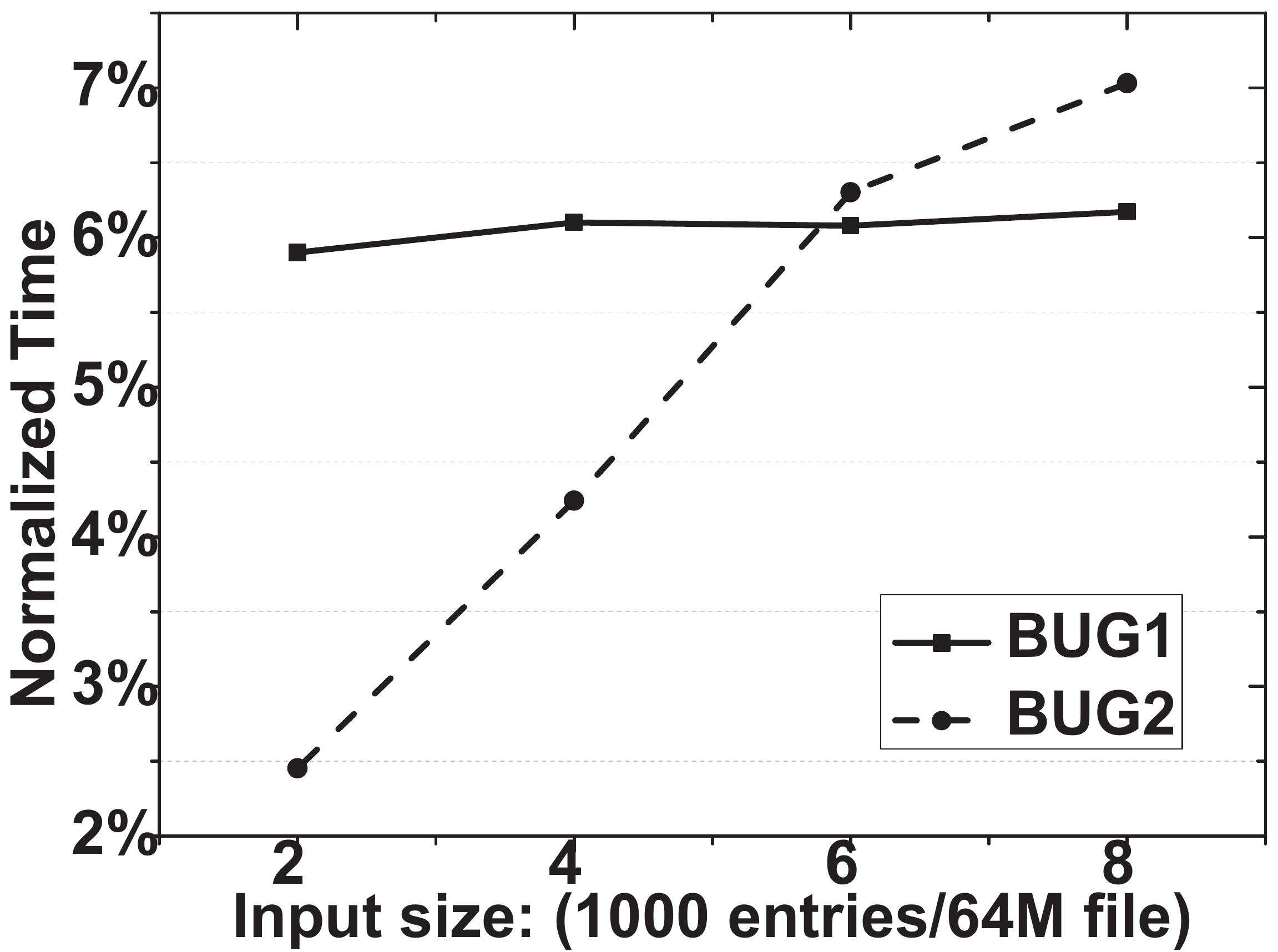}}
 \hfill
 \subfloat[Case study with the varying input size]{\includegraphics[scale=0.17]{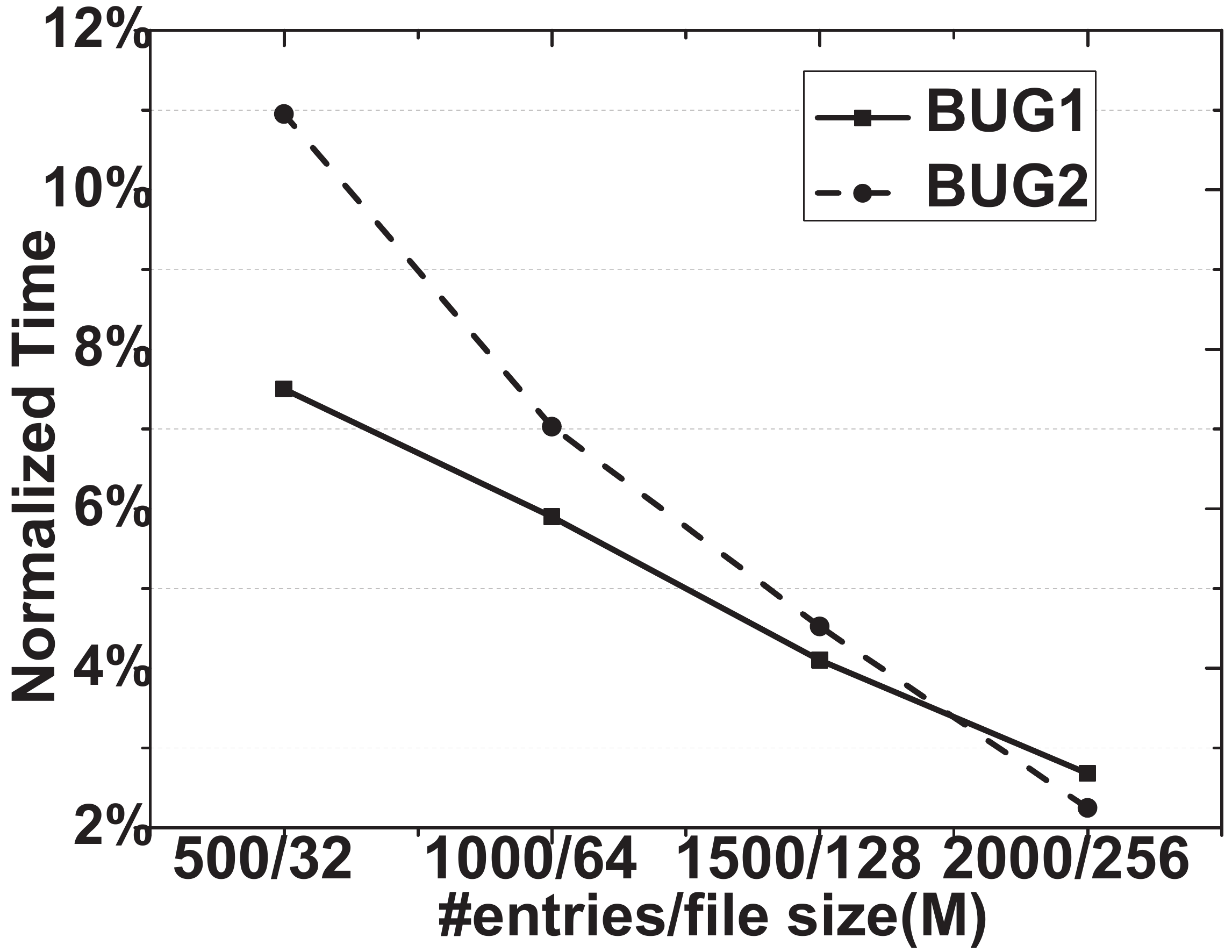}}
 \caption{Sensitivity study of \#BUG 1 and \#BUG 2}
 \label{fig:case:study}
\end{figure}

\subsection{Case Study}
\label{sec:case:study}
To evaluate the effectiveness of \textsc{PerfPlay}, we have checked some ULCP bugs that have already been verified by official bug system under \textsc{PerfPlay} framework.

{\bf MySQL \#68573}. Figure~\ref{fig:official:case} depicts the code snippet of this case from \textsf{mysql} version-$5.6.11$. The real designed intention of the programmers is that "a 50ms timeout for a SELECT statement waiting for the query cache lock is set. If the timeout expires, the statement executes without using the query cache"---\textsf{mysql} official documents. However, the ULCP performance problem "increases" this timeout threshold unwittingly when multiple threads invoke this code, thus severely degrading the efficiency of SELECT statement. Other examples in \textsf{mysql} include \textbf{\#37844}, \textbf{\#60951} and \textbf{\#69276}.

We also re-implement a few \emph{easy-to-understand} ULCP cases found by \textsc{PerfPlay} in a ULCP-free fashion, and further re-quantify its performance impact.

\textbf{Resource wasting from openldap} {\bf (\#BUG 1).} We re-implement the code snippet in Figure~\ref{fig:mtexample} with \texttt{pthread\_mutex\_barrier}, and re-quantify the CPU utilization of this ULCP problem by testing the program compared with the original code. 

\textbf{Performance degradation from pbzip2} {\bf (\#BUG 2).} Figure~\ref{fig:pbzip} depicts the simplified code of ULCP problem from the parallel compression utility pbzip2. It employs the producer-consumer idiom for the parallel compression: the producer produces the blocks by reading file and the multiple consumers consume (compress) these blocks in parallel. When the last file block is dequeued (i.e., \texttt{fifo->empty}=1 and \texttt{producerDone}=1), the program starts the end stage of thread join. In this case, the example above will incur many read-read ULCPs as follows:
\begin{lstlisting}[escapechar=@]
@\textbf{lock}@(mu);
@\textcolor[rgb]{0.25,0.00,0.50}{\textbf{load}}@(fifo->empty);
@\textbf{lock}@(muDone);@\textcolor[rgb]{0.25,0.00,0.50}{\textbf{load}}@(producerDone);@\textbf{unlock}@(muDone);
@\textbf{unlock}@(mu);
\end{lstlisting}
The joins of all threads are serialized and extra nested lock overhead is added by this read-read ULCP, which causes the performance loss. We fix it via the signal/wait model: we take the producer, rather than the consumer, with the responsibility of checking the state of \texttt{fifo->empty} and \texttt{producerDone}. If both of them are \textsf{true}, the producer will give a signal to inform all consumers of their safe exit without any check when their work is completed.

\textbf{Results}. Figure~\ref{fig:case:study} depicts the sensitivity of two exploited ULCP bugs (i.e., \#BUG 1 and \#BUG 2). As the number of threads increases, \#BUG 1 causes the stable resource wasting per thread while \#BUG 2 has an increasing performance loss of program. Whereas, different from the illustration shown in Figure~\ref{fig:perfImpact:input}(b), the performance impact of both \#BUG 1 and \#BUG 2 presents a downward trend as the input sizes increases. That is because for a given thread number both \#BUG 1 and \#BUG 2 have the fixed execution frequency, which rises superior to the input size. Moreover, the increasing input size aggravates the workload of application, thus increasing the program execution time. As a result, the performance impact of both \#BUG 1 and \#BUG 2 is declining. Both results verify that the real ULCPs can be exploited by \textsc{PerfPlay}.
\subsection{Discussion}
It is well-known that replay technique generally incurs the prohibitive overhead for the program execution. Hence it comes a big question that "\emph{is it reliable to use replay technique for the performance analysis?}" In our work, we investigate every time-consuming process of replay execution, such as the loading of trace from disk into memory, and the format transformation of trace. In the replay phase, we differentiate them from the real execution. Besides, in the real execution, there may be different strategies for the replay executions (such as, MEM-S and SYNC-S), which slows down the program execution to some extent. ELSC-S strategy eliminates the imprecise performance caused by them, further providing the faithfully original scheduling. Consequently, we argue that \textsc{PerfPlay} with ELSC-S strategy provides the correct ULCP recommendations although PIN framework may introduce the fairly-low (almost negligible~\cite{PIN}) baseline overhead. Our above-depicted results (shown in Section~\ref{sec:perf} and \ref{sec:case:study}) also verify this conclusion.

\textsc{PerfPlay} adapts one execution trace of the program to expose ULCPs, but we still believe it is useful to the programmers. First, we believe that performance debugging (ULCP in particular for this paper) is similar to the debugging process of common software bugs (e.g., error, failure, or fault). Programmers usually define a set of debugging cases. They usually perform step-by-step debugging as the program runs for a given input. This is simple and effective strategy. Likewise, we follow this strategy. Second, \textsc{PerfPlay}, still, can be extended to multiple traces.

\section{Related Work}
\label{sec:relatedwork}
\subsection{Unnecessary Lock Contention} 
There has been significant amount of work on dynamically eliminating the performance impact of ULCPs.
Lock Elision (LE)~\cite{SLE, Roy:2009} leverages the hardware assistance and the underlying cache coherence protocol to enable highly concurrent multi-threaded execution by dynamically removing unnecessary lock-induced serialization. The lock is acquired only when a data conflict occurs. However, LE-based work is still challenging in practice.
For instance, a few transaction aborts may cause excessive rollbacks and serializations, which severely limits the exposed concurrency of ULCPs \cite{Yehuda:2014}. Meanwhile, it is prone to trigger false aborts due to the hardware limitations \cite{Leis:2014}.
We believe that the most effective and efficient manner for ULCPs is that programmers can fix the problem in their code, rather than rely on dynamic tools which may lead to severe runtime overhead. Consequently, we propose a novel framework, \textsc{PerfPlay}, to evaluate the performance impact of ULCPs and further assist the programmers to identify the most performance critical ULCP.

\subsection{Performance Tools}
It is hard for static exploration tools \cite{Cadar:2011} to obtain the characteristics of ULCPs (e.g., their amounts, categories and the time they cost). Due to the dynamic nature of ULCPs, the major obstacle is that they may produce abundant false ULCPs due to the runtime behaviors of ULCPs. Another obstacle is that the code snippet with a lock/unlock pair running simultaneously by multiple threads may unroll into two execution cases as ULCPs and TLCPs. Under different runtime (e.g., thread scheduling and input set), both ULCPs and TLCPs manifest themselves in different amounts and performance impact. As for the existing dynamic tools~\cite{callstack,stackmine}, they also bear some limitations in the impact analysis of ULCPs. Still, the majority of them are devoted to performance measurement, but they are not applicable to the performance transformation and further performance comparison before and after optimization. As a result, they cannot be used directly for performance debugging, e.g., how much performance would be improved if the ULCPs were removed. \textsc{PerfPlay} is the very performance tool to make an attempt to solve this problem.

\subsection{Record/Replay System} 
Plentiful replay systems are proposed in the past several decades. For instance, deterministic replay systems \cite{pinplay, BugNet} reproduce the bug debugging by enforcing the order of the execution events. Modified replay debugging~\cite{Narayanasamy2007, MutableReplay} distinguishes different categories of bugs by comparing the results of the original trace with the modified one. Overall, almost all of them are built for identifying and understanding the correctness of bugs in programs. but not much effort has gone into the study of performance issues. \textsc{PerfPlay} first (to our best knowledge) has put effort into studying the performance bugs using replay technique.
\section{Conclusion and Future Work}
\label{sec:conclusion}
We propose a performance debugging framework, \textsc{PerfPlay}, to evaluate the performance impact of unnecessary lock contention pairs (ULCPs) in the multi-threaded applications. 
We first record the multi-threaded program execution trace, based on which we can identify all ULCPs. Then \textsc{PerfPlay} transforms the original ULCP trace into the new ULCP free one  while ascertaining the correctness of program via our four novel rules proposed. Finally, \textsc{PerfPlay} replays two traces. Based on two replayed results, we evaluate the potential performance improvement of each ULCP and then group all ULCPs into the unique ULCPs according to their code-site. Our experimental results on $5$ real world programs and PARSEC benchmarks demonstrate the performance fidelity and efficiency ($<4.3\%$ lockset overhead) of \textsc{PerfPlay}. With case studies, we demonstrate its effectiveness to identify the performance critical ULCP. It also shows that the majority of ULCPs can be resolved by taking the most critical code regions. We will make \textsc{PerfPlay} as a pintool in the PIN framework in the future.

\textsc{PerfPlay} currently helps the ULCP debugging of the input which produces that trace, but may not help the execution of program on other inputs. As a result, input sensitivity will give a great chance for us to make \textsc{PerfPlay} more useful, because this may prohibit any code modification that could lead to performance improvement in some cases but not all. We also plan to investigate the applicability of our tool to many-core programs (such as GPU-based applications~\cite{He:2008})

\section*{Acknowledgements}
The authors would like to thank the anonymous reviewers for their valuable comments. Special thanks also go to Xuelin Hu and Chencheng Ye who help us a lot to improve and perfect this work selflessly.



\bibliographystyle{abbrv}
\bibliography{reference}

\section*{Appendix A: Cases in the Real World}
\label{appendix}
We list some real ULCP cases mainly used for the discussion and understanding of ULCP manifestation. Some of them may not introduce the serious performance impact. Whilst some come from the officially verified ULCP bugs, which lead to the performance problem of program.
\begin{figure}[h] 
\centering
\includegraphics[scale=0.85]{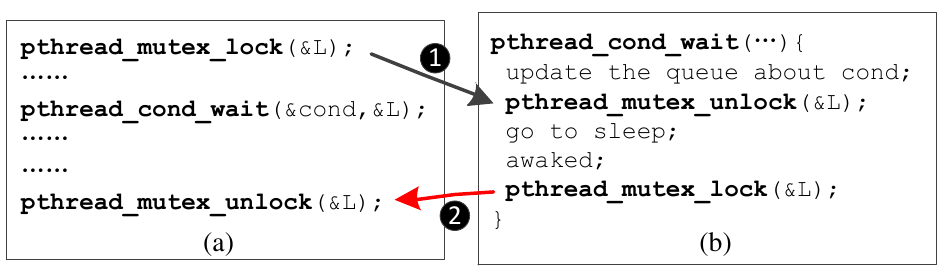}
\caption{Case 1}
\label{fig:0}
\end{figure}

{\bf Case 1:} This case shows how the typical implementation of \texttt{pthread\_cond\_wait} in \texttt{pthread} library contributes to the ULCP. (a) In order to protect the correct update on \texttt{cond}, \texttt{pthread\_cond\_wait} is usually used within a lock/unlock pair. (b) Such usage may lead to ULCP in the current implementation of \texttt{pthread\_cond\_wait}, which first releases the lock before the thread goes to sleep, and then re-acquires the lock before it exits. As a result, the second lock/unlock pair may lead to a NULL-Lock where no shared data is accessed.

\begin{figure}[h] 
\centering
\includegraphics[scale=1]{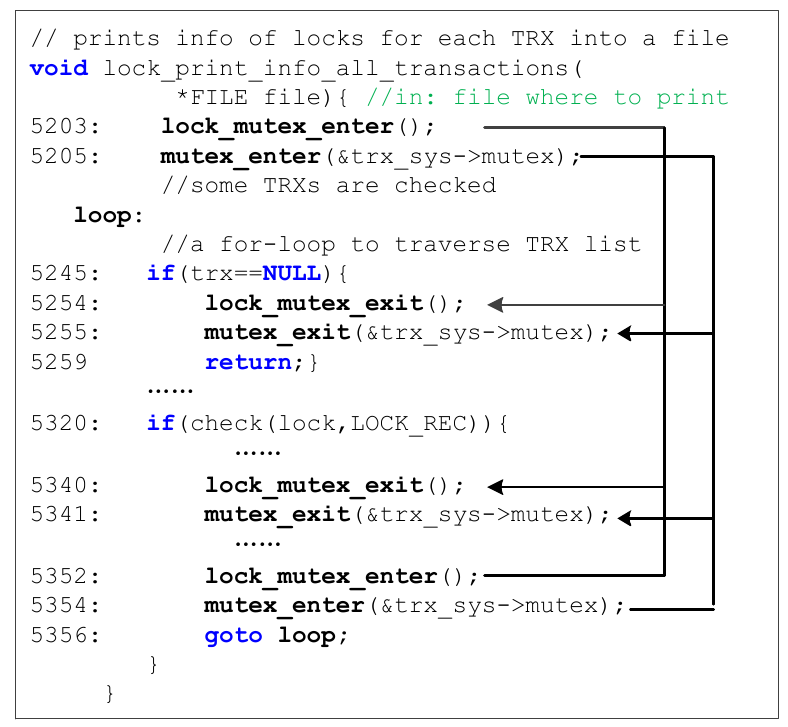}
\caption{Case 2}
\label{fig:2}
\end{figure}

{\bf Case 2:} This code snippet in Figure~\ref{fig:2} has traversed the entire transaction (TRX) list to print the information of locks for each TRX into a given file. If the multiple threads invoke this code simultaneously, many read-read pattern of ULCPs occur.

\begin{figure}[tbph] 
\centering
\includegraphics[scale=1.3]{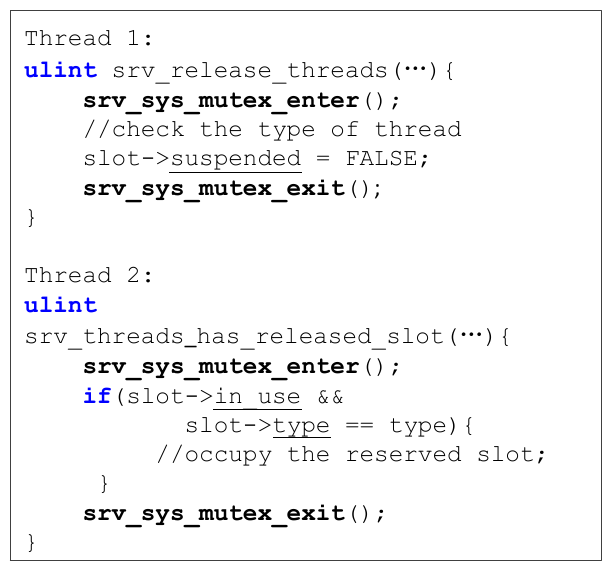}
\caption{Case 3}
\label{fig:3}
\end{figure}

{\bf Case 3:} \texttt{slot->suspended} is updated by Thread 1, and Thread 2 only reads the value of \texttt{slot->in\_use} and \texttt{slot->type}. Both threads access the same shared object \texttt{slot}, but they do not essentially conflict due to the fact that the disjoint fields of \texttt{slot} are accessed.

\begin{figure}[tbp] 
\centering
\includegraphics[scale=1]{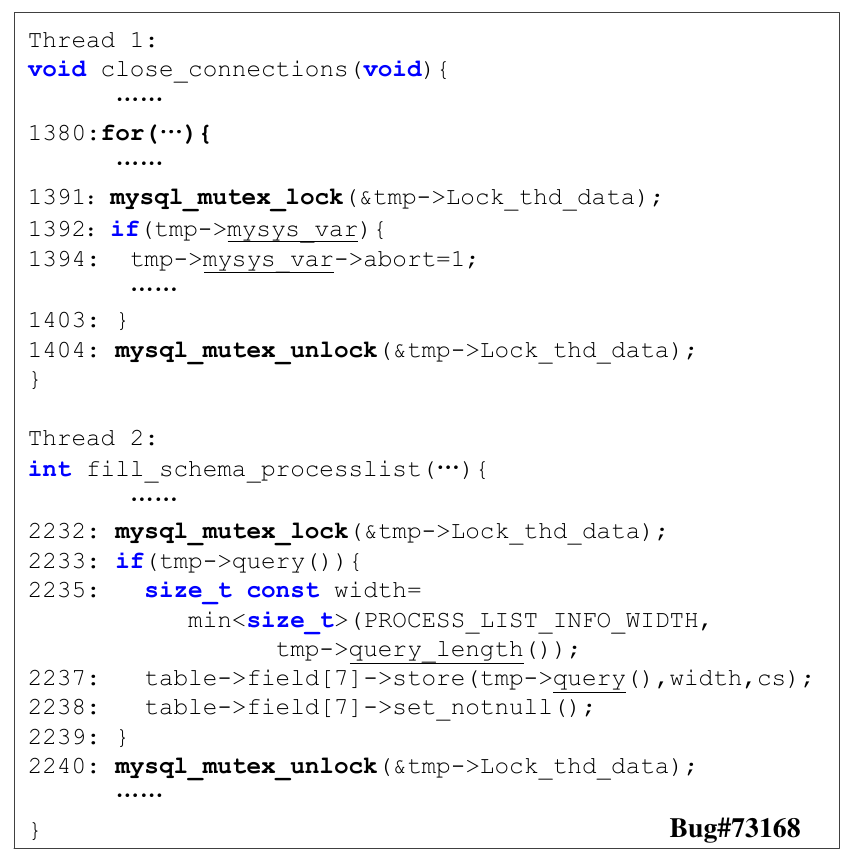}
\caption{Case 4}
\label{fig:4}
\end{figure}

{\bf Case 4:} For the shared lock \texttt{LOCK\_thd\_data}, it is major used to protect \texttt{thd->query} and \texttt{thd->query\_length}. In this example, \texttt{Lock\_thd\_data} is also used to protect \texttt{thd->mysys\_var} when the thread exits by invoking \texttt{close\_contections}, thus blocking the proceeding of the query manipulation.

\begin{figure}[tbp] 
\centering
\includegraphics[scale=1]{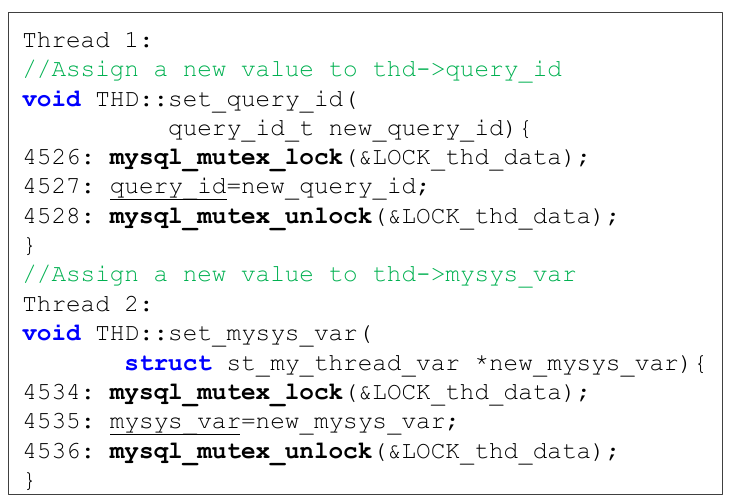}
\caption{Case 5}
\label{fig:5}
\end{figure}

{\bf Case 5:} both \texttt{THD::set\_query\_xxx()} use the same shared lock \texttt{LOCK\_thd\_data} to assign a new value to the different member, we can benefit with less overhead if replacing mutex with lock-free atomic operations.

\begin{figure}[tbp] 
\centering
\includegraphics[scale=1]{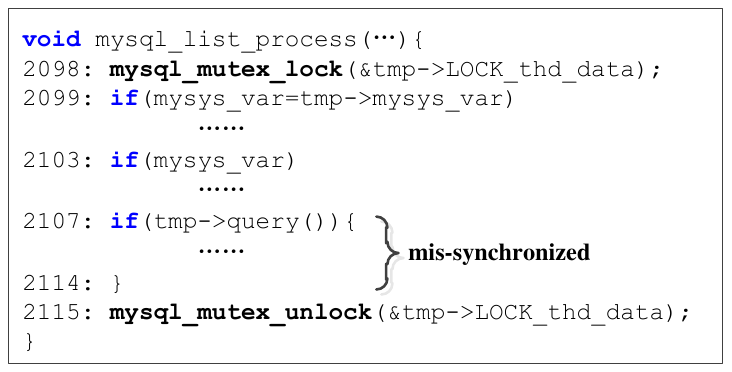}
\caption{Case 6}
\label{fig:6}
\end{figure}

{\bf Case 6:} coarse-grained lock synchronization is used to protect a large, complex transaction which, in fact, can be partitioned. 

\begin{figure}[tbp] 
\centering
\includegraphics[scale=1.1]{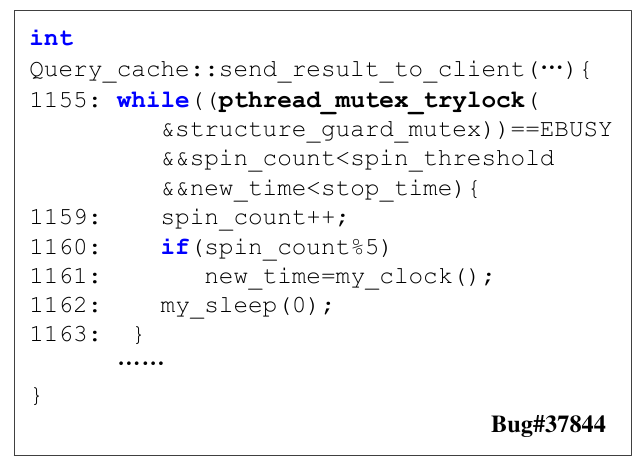}
\caption{Case 7}
\label{fig:7}
\end{figure}

{\bf Case 7:} only 1 thread can search the query cache at a time from mysql-5.0(bug \#37844). A pthread\_mutex is held while it is searched and this will limit performance on multi-core servers. Besides, this code snippet makes an attempt to use a spin lock, but the spin lock also wastes a large amount of CPU time (i.e., system throughput loss).

\begin{figure}[tbp] 
\centering
\includegraphics[scale=1.1]{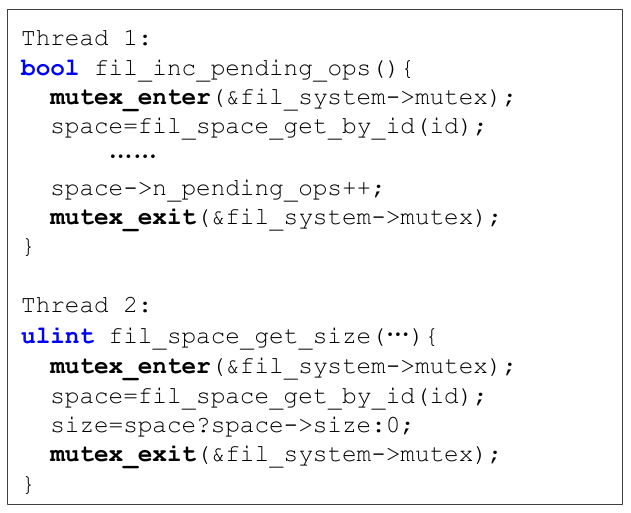}
\caption{Case 8}
\label{fig:8}
\end{figure}

{\bf Case 8:} For each block read operation, the hash table lookup \texttt{fil\_space\_get\_by\_id} is done at least four times by \texttt{fil\_space\_get\_version}, \texttt{fil\_inc\_pending\_ops}, \texttt{fil\_decr\_pending\_ops} and  \texttt{fil\_space\_get\_size} (MySQL bug \#69276). Suppose multiple threads proceed a large number of read-only transactions, this type of ULCP problem serializes all lookups of the hash table with a slowdown of 4X at least.

\begin{figure}[tbp] 
\centering
\includegraphics[scale=1.1]{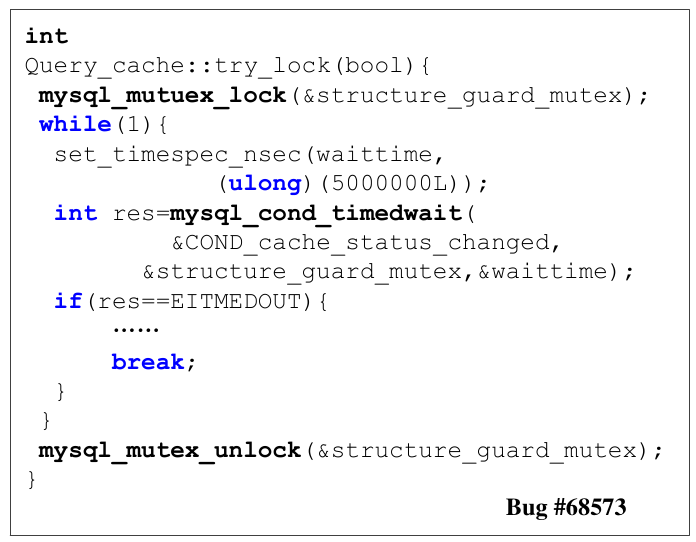}
\caption{Case 9}
\label{fig:9}
\end{figure}

\begin{figure}[h] 
\centering
\includegraphics[scale=0.8]{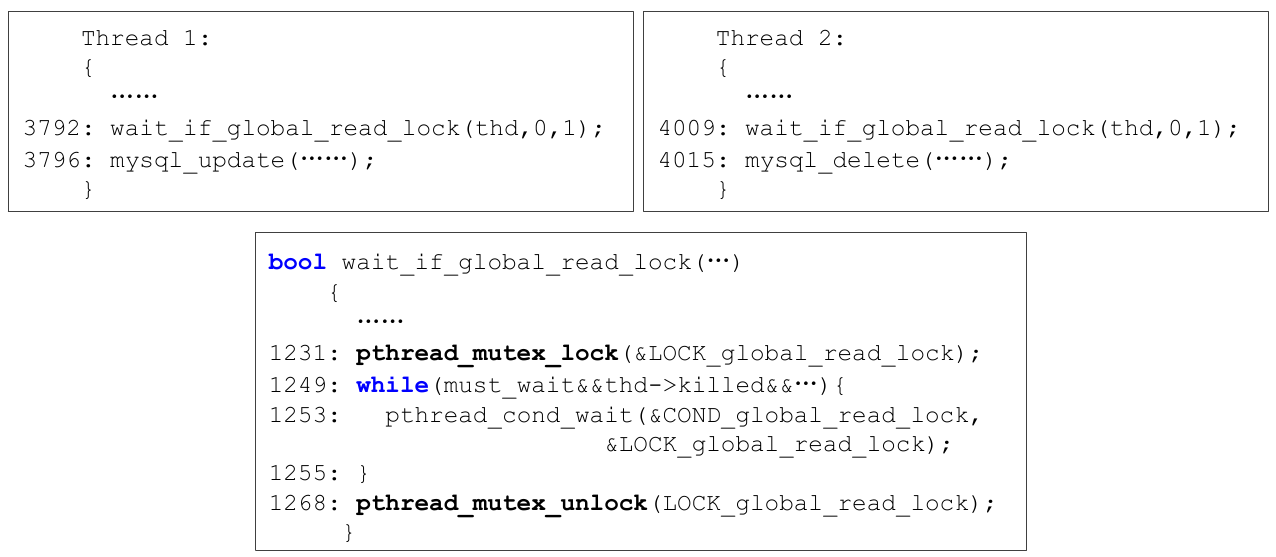}
\caption{Case 10}
\label{fig:10}
\end{figure}

{\bf Case 9:} This case comes from mysql version-5.6.11. The real designed intention of the programmers is that "a 50ms timeout for a SELECT statement waiting for the query cache lock is set. If the timeout expires, the statement executes without using the query cache" ---from \texttt{mysql} Documents. However, the ULCP performance problem "increases" this timeout threshold unwittingly, which severely degrades the efficiency of SELECT statement of \texttt{mysql}.

{\bf Case 10:} This case illustrates a complex ULCP performance pattern which serializes the UPDATE and SELECT statement even if they manipulate the different fields of the table (mysql bug \#60951).

More cases will be appended soon and available at http://211.69.198.173/longzh/ulcp.php/.

\end{document}